\documentclass[onecolumn,floatfix,superscriptaddress,showpacs,showkeys,nofootinbib,preprint]{revtex4}
\usepackage{epsfig}
\usepackage{amssymb,latexsym,amsmath}
\newcommand{\eq}[1]{\begin{align} #1 \end{align}}

\begin{document}

\title{Multiplicity Fluctuations in Nucleus-Nucleus Collisions: \\
Dependence on Energy and Atomic Number
}

\author{V.P.~Konchakovski}
\affiliation{Helmholtz Research School, University of Frankfurt, Frankfurt, Germany}
\affiliation{Bogolyubov Institute for Theoretical Physics, Kiev, Ukraine}
\author{B. Lungwitz}
\affiliation{Institut f\"ur Kernphysik, University of Frankfurt, Frankfurt, Germany}
\author{M.I.~Gorenstein}
\affiliation{Bogolyubov Institute for Theoretical Physics, Kiev, Ukraine}
\affiliation{Frankfurt Institute for Advanced Studies, Frankfurt, Germany}
\author{E.L.~Bratkovskaya}
\affiliation{Frankfurt Institute for Advanced Studies, Frankfurt, Germany}

\begin{abstract}
Event-by-event multiplicity fluctuations in central C+C, S+S,
In+In, and Pb+Pb as well as p+p collisions at bombarding energies
from 10 to 160~AGeV are studied within the HSD and UrQMD
microscopic transport approaches.  Our investigation is directly
related to the future experimental program of the NA61
Collaboration at the SPS for a search of the QCD critical point.
The dependence on energy and atomic mass  number of the scaled variances
for negative, positive, and all charged hadrons is presented and
compared to the results of the model of independent sources.
Furthermore, the nucleus-nucleus results from the transport
calculations are compared to inelastic proton-proton collisions
for reference.  We find a dominant role of the participant number
fluctuations in nucleus-nucleus reactions at finite impact
parameter $b$. In order to reduce the influence of the participant
numbers fluctuations on the charged particle multiplicity
fluctuations only the most central events have to be selected.
Accordingly, the samples of the 1\%  most central nucleus-nucleus
collisions with the largest numbers of the projectile participants
are studied. The results are compared with those for  collisions
at zero impact parameter. A strong influence of the centrality
selection criteria on the multiplicity fluctuations  is pointed
out. Our findings are essential for an optimal choice of colliding
nuclei and bombarding energies for the experimental search of the
QCD critical point.

\end{abstract}


\maketitle 

\section{Introduction \label{intr}}

The event-by-event fluctuations in high energy nucleus-nucleus
(A+A) collisions (see e.g., the reviews \cite{rev1,rev2,rev3}) are
expected to be closely related to the transitions between
different phases of QCD matter. By measuring the fluctuations one
should observe  anomalies from the onset of deconfinement
\cite{ood} and dynamical instabilities when the expanding system
goes through the 1-st order transition line between the
quark-gluon plasma (QGP) and the hadron gas \cite{fluc2}.
Furthermore, the QCD critical point may be signaled by a
characteristic pattern in enhanced fluctuations.  A+A collisions
in the SPS energy region are expected to be a suitable tool for a
search of critical point signatures \cite{SRS}.  Only recently
first measurements of particle multiplicity fluctuations
\cite{fluc-mult} and transverse momentum fluctuations
\cite{fluc-pT} in A+A collisions have been performed. A
theoretical analysis of multiplicity fluctuations for the
hadron-resonance gas - in different statistical ensembles - has
been performed in Ref.~\cite{CE}. Independently, the multiplicity
fluctuations in A+A collisions has been studied within a
microscopic transport approach \cite{KGB1,KGB2}.  We recall that
fluctuations traditionally are quantified by the ratio of the
variance of the multiplicity distribution to its mean value, the
scaled variance. Previous works on this subject have to be quoted.
The calculations of the statistical models \cite{MCE} and
transport approaches HSD \cite{KGB3} and UrQMD v. 1.3 \cite{LB}
have been compared with the corresponding preliminary results
\cite{NA49} of the NA49 Collaboration in central Pb+Pb collisions
at  SPS energies. At RHIC energies a HSD analysis of the
preliminary data \cite{phenix} of the PHENIX Collaboration in
Au+Au collisions at $\sqrt{s_{NN}}=200$~GeV has been presented in
Ref.~\cite{KGB4}.

An ambitious experimental program for the search of the QCD
critical point has been started by the NA61 Collaboration  at the
SPS \cite{NA61}. The program includes a variation in the  atomic mass 
number $A$ of the colliding nuclei as well as an energy scan. This
allows to scan the phase diagram in the plane of temperature $T$
and baryon chemical potential $\mu_B$ near the critical point as
argued in Ref.~\cite{NA61}. One expects to `locate' the position
of the critical point by studying its `fluctuation signals'. High
statistics multiplicity fluctuation data will be taken for p+p,
C+C, S+S, In+In, and Pb+Pb collisions at bombarding energies of
$E_{lab}$=10, 20, 30, 40, 80, and  158~AGeV.

The aim of the present paper is to study the energy and system
size dependence of  event-by-event multiplicity fluctuations
within the microscopic transport approaches Hadron-String-Dynamics
(HSD, v. 2.5) \cite{HSD} and
Ultra-Relativistic-Quantum-Molecular-Dynamics (UrQMD, v 1.3)
\cite{UrQMD}. These models provide a rather reliable description
(see, e.g., Refs. \cite{Weber,UrQMD}) for the inclusive spectra of
charged hadrons in A+A collisions from SIS to RHIC energies. In
our study we will consider C+C, S+S, In+In, and Pb+Pb collisions
at the bombarding energies of  10,~ 20,~ 30,~ 40,~ 80,~ 158~AGeV.
For a comparison  and reference we also present the results of
multiplicity fluctuations in p+p collisions at the same energies.
Our study thus is in full correspondence to the experimental
program of the NA61 Collaboration \cite{NA61}. In order to
estimate systematical errors from the theoretical side we employ
the two independent transport models (HSD and UrQMD) as in Ref.
\cite{Weber}.

The QCD critical point is expected to be experimentally seen as a
non-monotonic dependence of the multiplicity fluctuations, i.e. a
specific combination of atomic mass  number $A$ and bombarding energy
$E_{lab}$ could move the chemical freeze-out of the system close
to the critical point and show a `spike' in the multiplicity
fluctuations. Since  HSD and UrQMD do not include explicitly a
phase transition from a hadronic to a partonic phase, we can not
make a clear suggestion for the  location of the critical point -
it is beyond the scope of our hadron-string models.  However, our
study might be helpful in the interpretation of the upcoming
experimental data since it will allow to subtract simple dynamical
and geometrical effects from the expected QGP signal.  The
deviations of the future experimental data from the HSD and UrQMD
predictions may be considered as an indication for the critical
point signals.

Theoretical estimates give about 10\%  increase of the
multiplicity fluctuations due to the critical point \cite{SRS}. It
is large enough to be observed experimentally within the
statistics of NA61 \cite{NA61}. To achieve this goal, it is
necessary to have a control on other possible sources of
fluctuations. One of such sources is the fluctuation of the number
of nucleon participants. It has been found in Ref. \cite{KGB1}
that these fluctuations give a dominant contribution to hadron
multiplicity fluctuations in A+A collisions. On the other hand one
can suppress the participant number fluctuations by selecting most
central A+A collisions (see Ref.~\cite{KGB1,MGMG} for details).
That's why the NA61 Collaboration plans to measure central
collisions of light and intermediate ions instead of peripheral
Pb+Pb collisions.  It is important to stress, that the conditions
for the centrality selection in the measurement of fluctuations
are much more stringent than those for mean multiplicity
measurements. This issue will be discussed in detail in our paper.

Our paper is organized as follows. In Section II the HSD and UrQMD
models are compared with the data for the charged hadron
multiplicity, mean values and fluctuations in p+p collisions in
the SPS energy range $10-158$~AGeV.  In Section III the
participant number fluctuations in fixed target experiments are
discussed and estimated within  HSD and UrQMD. In Section IV we
present the HSD and UrQMD results for the charged hadron
multiplicity fluctuations in $A+A$ collisions with zero impact
parameter, $b=0$. The participant number fluctuations in $A+A$
collisions at $b=0$ are considered  in Section V. In addition, the
transport model results for the charged hadron multiplicity
fluctuations are compared within the model of independent sources.
In Section VI the centrality selection - by fixing the number of
projectile participants - is considered  and compared to the case
of $b=0$.  Section VI presents also the HSD and UrQMD results for
the charged hadron multiplicity fluctuations in $A+A$ collisions
for the 1\%  most central collisions corresponding to the largest
number of projectile participants.  A summary and conclusion will
close the paper in Section VII.

\section{Multiplicity fluctuations in proton-proton
collisions}\label{pp}

For a quantitative measure of the particle number fluctuations it is
convenient to use the scaled variances,
\eq{
 \omega_i~\equiv~\frac{\langle N_i^2\rangle~-~\langle
N_i\rangle^2}{\langle N_i\rangle}~,
 \label{omega_def}
}
where $\langle \cdots\rangle$ denotes event-by-event averaging and
the index $i$ means ``-'', ``+'', and ``ch'', i.e negative,
positive, and all charged final state hadrons.

The energy dependence of the measured charged multiplicity and
fluctuations for p+p collisions can be parametrized by the
functions \cite{rev1}:

\eq{\label{Nchpp}
\langle N_{ch}\rangle~\cong~
-4.2~+~4.69~\left(\frac{\sqrt{s_{NN}}}{\mbox{GeV}}\right)^{0.31}~,~~~~
 \omega_{ch}~\cong~ 0.35~\frac{(\langle
N_{ch}\rangle~-~1)^2}{\langle N_{ch}\rangle}~,
}
where $\sqrt{s_{NN}}$ is the center-of-mass energy.

\begin{figure}[ht!]
\epsfig{file=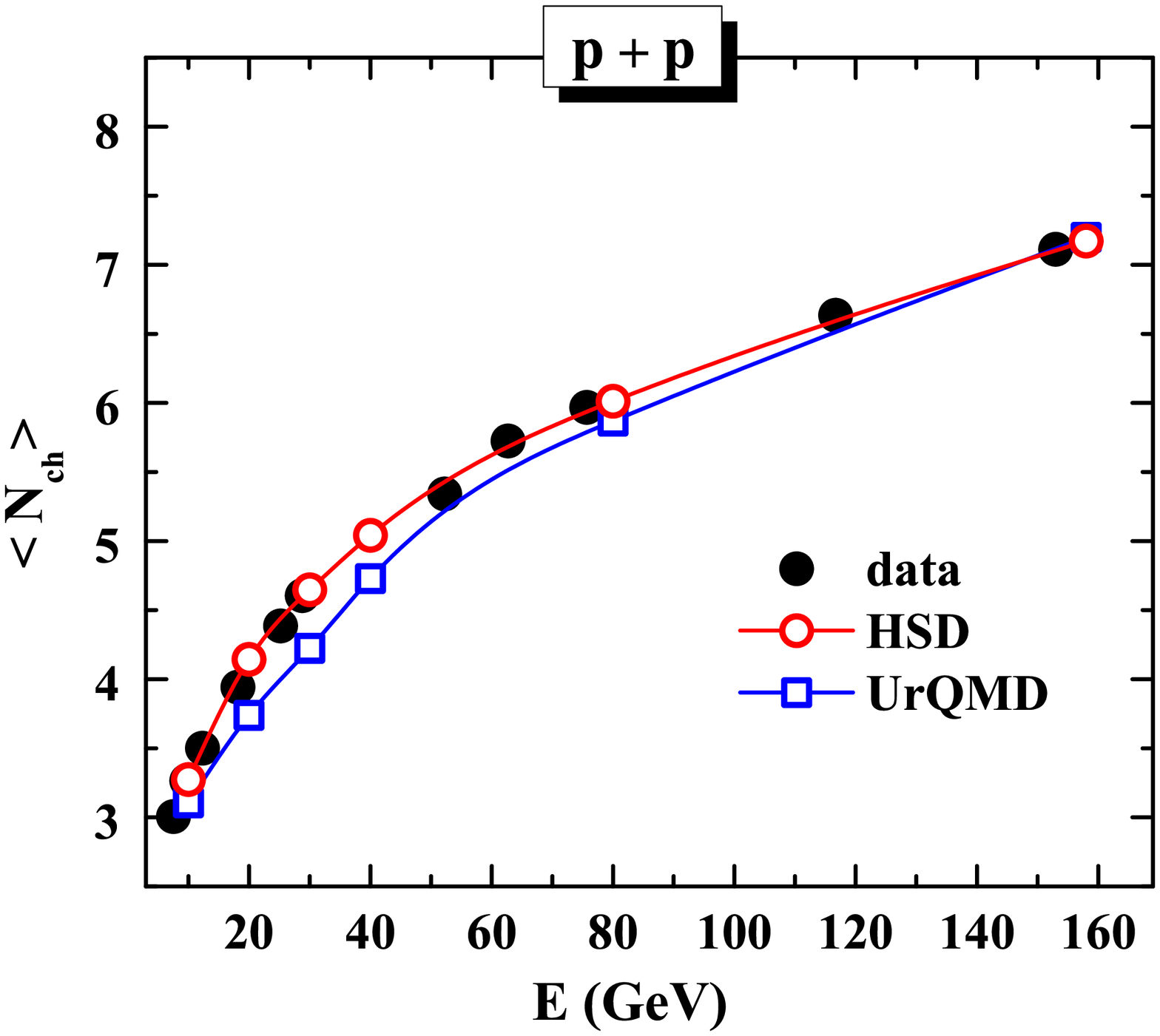,height=7cm}
\epsfig{file=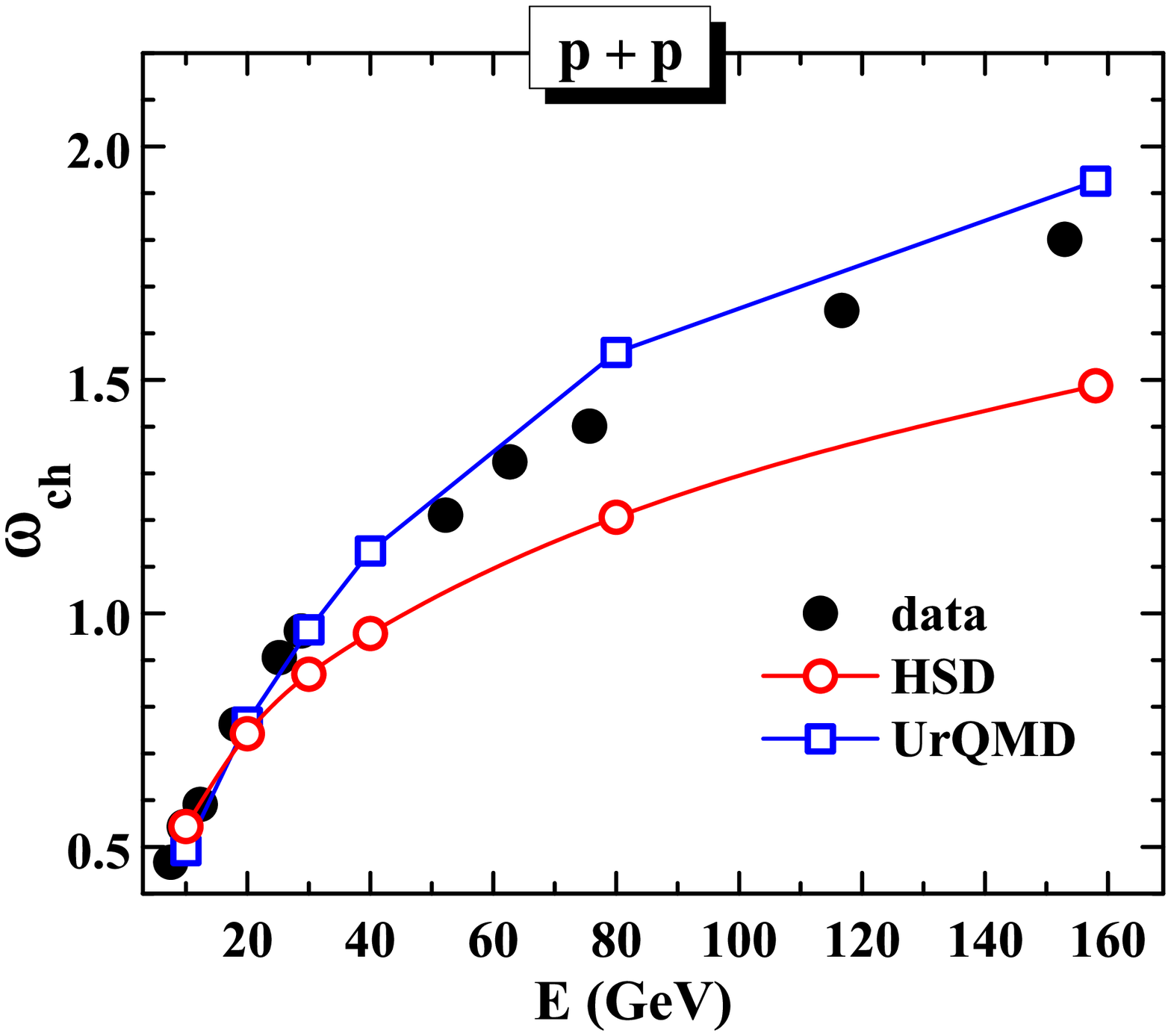,height=7cm}
\caption{(Color online) The average multiplicity ({\it left}) and scaled
variance ({\it right}) of charged hadrons in p+p inelastic
collisions. The open circles and squares (connected by solid
lines) show the results of HSD and UrQMD, respectively, whereas
the full circles present the experimental data from
Ref.~\cite{rev1}.} \label{flucP}
\end{figure}

Figure \ref{flucP} shows the HSD and UrQMD results for  inelastic p+p
collisions in comparison to  the experimental data taken from
Ref.~\cite{rev1}.  As seen from Fig. \ref{flucP} both models give a
good reproduction of the p+p data for $\langle N_{ch}\rangle$, but
slightly (over) underestimate $\omega_{ch}$ at high collision energies.
The differences between the HSD and UrQMD model results for
$\omega_{ch}$ can be attributed to different realizations of the string
fragmentation model, in particular,  differences in the fragmentation
functions and the fragmentation scheme, i.e. fragmentation via heavy
baryonic and mesonic resonances in UrQMD and direct light hadron
production by string fragmentation in HSD.

For negative and positive charged hadrons the average multiplicities
and scaled variances  in p+p collisions can be presented in terms
of the corresponding quantities for all charged particles,
\eq{ \label{posneg}
\langle N_{\pm}\rangle~=~\frac{1}{2}~ \left(\langle
N_{ch}\rangle~\pm~ 2\right)~,~~~~ \omega_{\pm}~=~\frac{1}{2}~
\omega_{ch} ~\frac{\langle N_{ch}\rangle}{\langle
N_{ch}\rangle~\pm~ 2}~.
}

\section{Participant number fluctuations \label{Np-fluc}}

In each $A+A$ collision only a fraction of all 2$A$ nucleons,
i.e. the participants, interact. Participants from the projectile
and target nuclei are denoted as $N_P^{proj}$ and $N_P^{targ}$,
respectively. The nucleons, which do not interact, are denoted as
projectile and target spectators, $N_S^{proj} = A - N_P^{proj}$
and $N_S^{targ} = A - N_P^{targ}$. The spectators are selected
according to the criteria $|y-y_{beam(target)}|\le 0.32$
and excluded from the multiplicity fluctuation analysis.
We recall that fluctuations in high energy $A+A$ collisions are
dominated by a geometrical  variation of the impact parameter $b$.
However, even for  fixed impact parameter $b$ the number of
participants, $N_P\equiv N_P^{proj}+N_P^{targ}$, fluctuates from event
to event.  This is due to fluctuations in the initial states of the
colliding nuclei and the probabilistic character of the interaction
process. These fluctuations of $N_P$ form usually a large and
uninteresting background.

To minimize the event-by-event fluctuations of the number of
nucleon participants in measuring the multiplicity fluctuations
the NA49 Collaboration has been trying to fix $N_P^{proj}$ in Pb+Pb
collisions. Samples of collisions with a fixed number of
projectile spectators, $N_S^{proj} = const$ (and thus a fixed
number of projectile participants, $N_P^{proj}$), have been selected.
This selection is possible in fixed target experiments at the SPS,
where $N_S^{proj}$ is measured by a Zero Degree Veto Calorimeter
covering the projectile fragmentation domain. A similar
centrality selection is expected to be implemented in the future
NA61 experiment. However, even in  samples with $N_P^{proj} =
const$ the number of target participants will fluctuate
considerably. Hence, an asymmetry between projectile and target
participants is introduced, i.e. $N_P^{proj}$ is constant by
constraint, whereas $N_P^{targ}$ fluctuates independently (the
consequences of this asymmetry have been discussed in Ref.~\cite{MGMG}).

In each sample with $N_P^{proj}=const$ the number of target
participants fluctuates around its mean value with the scaled variance
$\omega_P^{targ}$. The mean value equals to $\langle N_P^{targ} \rangle
\cong N_P^{proj}$, if $N_P^{proj}$ is not too close to its limiting
values, $N_P^{proj}=1$ and $N_P^{proj}$=A.  The scaled variance of
target participants $\omega_P^{targ}$ as a function of fixed number of
projectile participants $N_P^{proj}$ has been obtained from  HSD and
UrQMD in Ref.~\cite{KGB1} for Pb+Pb collisions at 158~AGeV.

\begin{figure}[ht!]
\epsfig{file=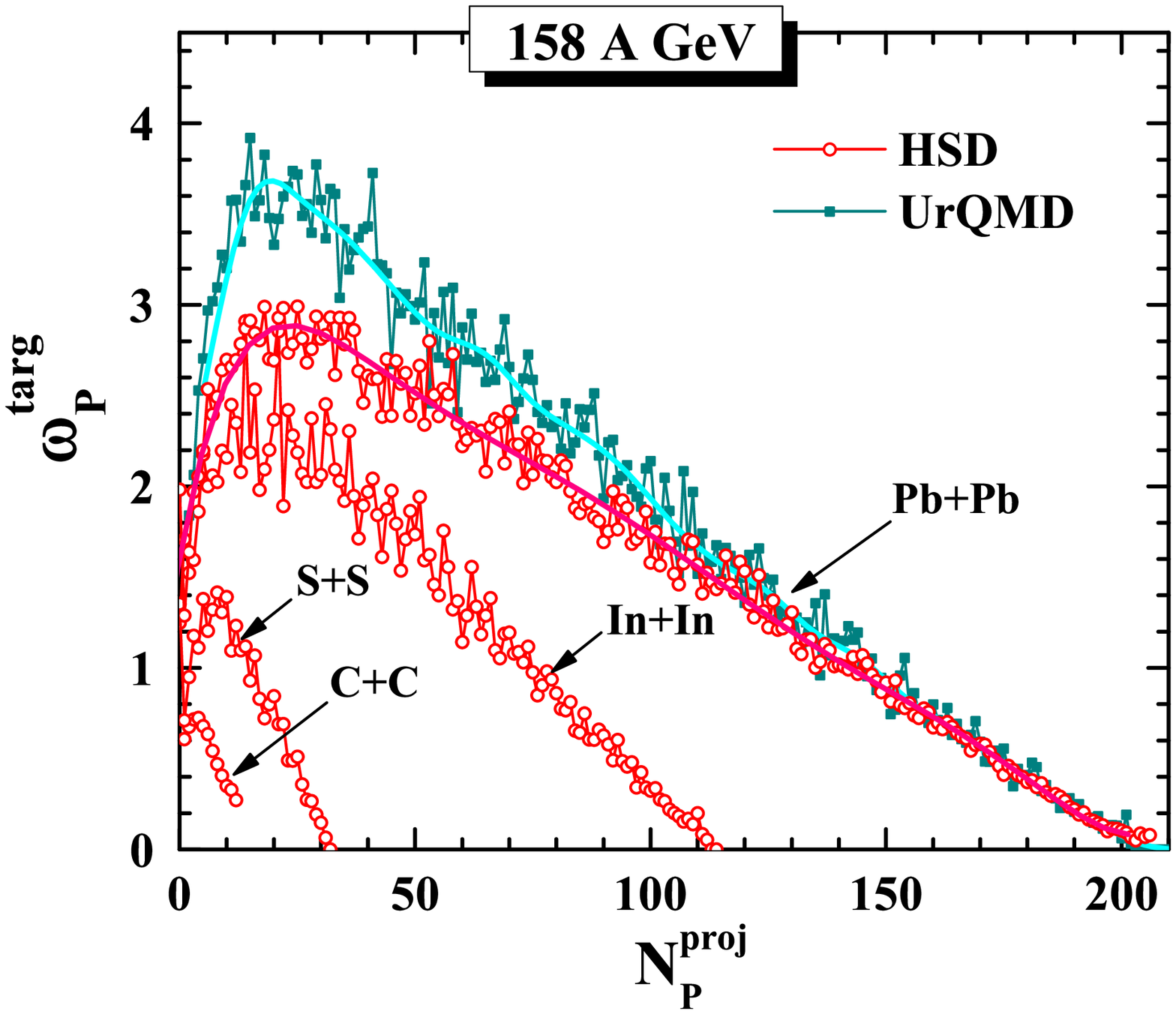,height=7cm}
\epsfig{file=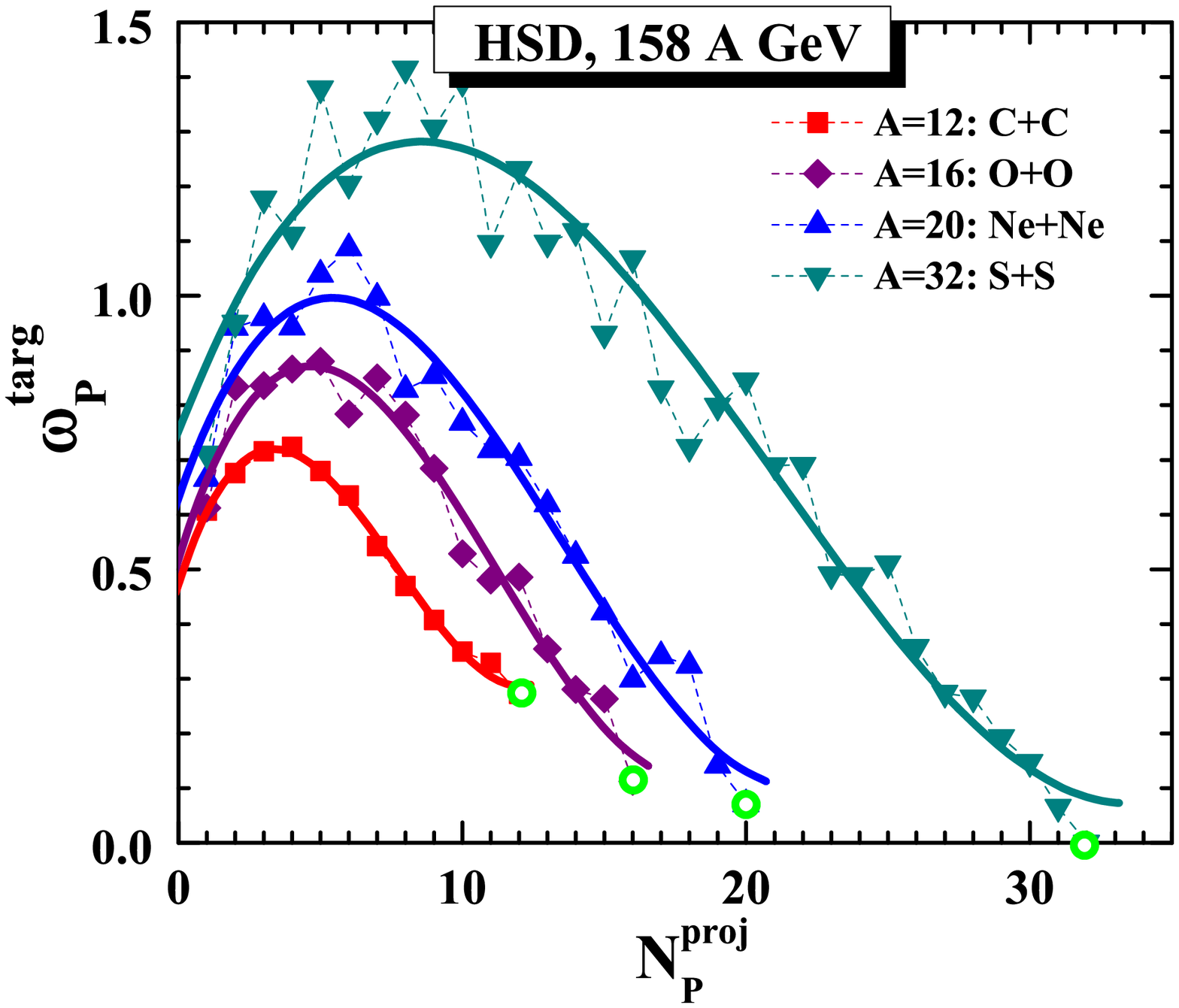,height=7cm} \caption{(Color online) {\it Left: } The
scaled variance $\omega_P^{targ}$  for the fluctuations of the
number of target participants, $N_P^{targ}$. The HSD simulations
of $\omega_P^{targ}$ as a function of $N_P^{proj}$ are shown for
different colliding nuclei, In+In, S+S, and C+C at
$E_{lab}$=158~AGeV. The HSD and UrQMD results for Pb+Pb collisions
are taken from Ref.~\cite{KGB1}. {\it Right:} The scaled variance
$\omega_P^{targ}$ for light nuclei, S+S, Ne+Ne, O+O, and C+C at
$E_{lab}$=158~AGeV as a function of $N_P^{proj}$. Fluctuations of
target participants for $N_P^{proj}=$A - shown by the open symbols
- are different from zero and increase with decreasing atomic mass 
number $A$. } \label{w_targ}
\end{figure}

Fig.~\ref{w_targ} ({\it left}) presents the HSD scaled variances
$\omega_P^{targ}$ for C+C, S+S, In+In, and Pb+Pb collisions at
158~AGeV as a function of $N_P^{proj}$. The fluctuations of
$N_P^{targ}$ are quite strong for peripheral reactions (small
$N_P^{proj}$)  and negligible for the most central collisions
(large $N_P^{proj}$). A vanishing of $\omega_P^{targ}\cong 0$ at
$N_P^{proj}\cong $A does not, however, show up  in C+C collisions.
Even for $N_P^{proj}$=A=12 in C+C collisions the number of
participants from the target still fluctuates and the scaled
variance amounts to $\omega_P^{targ}\cong 0.25$. Fig.~\ref{w_targ}
({\it right}) shows $\omega_P^{targ}$ for light nuclei, S+S,
Ne+Ne, O+O, and C+C. Even for the  maximal values of
$N_P^{proj}=$A the fluctuations $\omega_P^{targ}$ do not vanish
and increase with decreasing atomic mass  number $A$.

The temperature $T$ and baryon chemical potential $\mu_B$ at the
hadron chemical freeze-out demonstrate the dependence on both the
collision energy and system size \cite{Becattini}. Thus, changing
the number of participating nucleons one may scan the $T-\mu_B$
plane. Some combination of $N_P$ and $E_{lab}$ might move the
chemical freeze-out point close the QCD critical point. One could
then expect an increase of multiplicity fluctuations in comparison
to their `background values'.

Why does one need central collisions of light and intermediate
ions instead of studying peripheral Pb+Pb collisions for a search
of the critical point?  Fig.~\ref{w_targ} explains this issue. At
fixed $N_P^{proj}$ the average total number of participants,
$N_P\equiv N_P^{proj}+N_P^{targ}$, is equal to $\langle N_P\rangle
\cong 2 N_P^{proj}$, and, thus, it fluctuates as
$\omega_P=0.5\omega^{targ}_P$. Then, for example, the value of
$N_P^{proj}\cong 30$ corresponds to almost zero participant number
fluctuations, $\omega_P\cong 0$, in S+S collisions while
$\omega_P$ becomes large and is close to 1 and 1.5 for In+In and
Pb+Pb, respectively. Even if $N_P^{proj}$ is fixed exactly, the
sample of the peripheral collision events in the heavy-ion case
contains large fluctuations of the participant number:  this would
`mask' the critical point signals. As also seen in
Fig.~\ref{w_targ} ({\it right}), the picture becomes actually more
complicated if the atomic mass  number $A$ is too small. In this case,
the number of participants from a target starts to fluctuate
significantly even for the largest and fixed value of
$N_P^{proj}=$A.

\section{Multiplicity fluctuations at zero impact parameter}

\subsection{Centrality Selection in A+A Collisions by
Impact Parameter}

The importance of a selection of the most central collisions for
studies of hadron multiplicity fluctuations has been stressed in
our previous papers \cite{KGB1,KGB2,KGB3,KGB4}. Due to its
convenience in theoretical studies (e.g., in hydrodynamical
models) one commonly uses the condition on impact parameter $b$,
for the selection of the  `most central' collisions in model
calculations. However, the number of participant even at $b=0$ is
not strictly fixed, and fluctuates according to some distributions
(cf. Fig. 14 ({\it right}.) from Ref. \cite{KGB2}). It should be
stressed again that the conditions $b<b_{max}$ can not be fixed
experimentally since the impact parameter itself can not be
measured in a straightforward way. Actually, in experiments one
accounts for the 1\%, 2\% etc. most central events selected by the
measurement of spectators in the Veto calorimeter, which
corresponds to the event class with the largest $N_P^{proj}$.  As
we will demonstrate below  the multiplicity fluctuations are very
sensitive to the centrality selection criteria. In particular, the
transport model results for $b=0$ and for 1\% events with the largest
$N_P^{proj}$ are rather different (see below).

\begin{figure}[ht!]
\epsfig{file=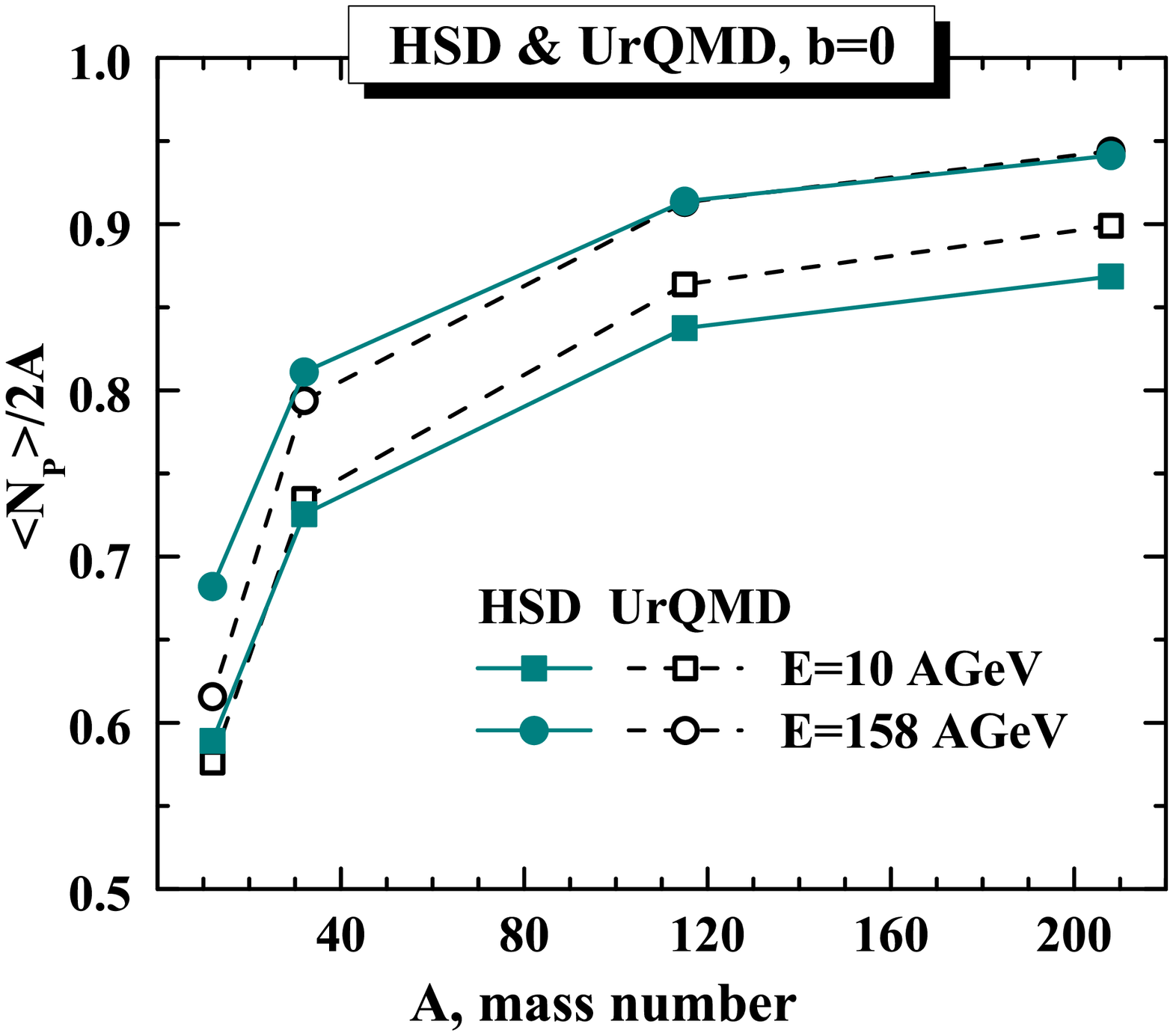,height=7cm}\epsfig{file=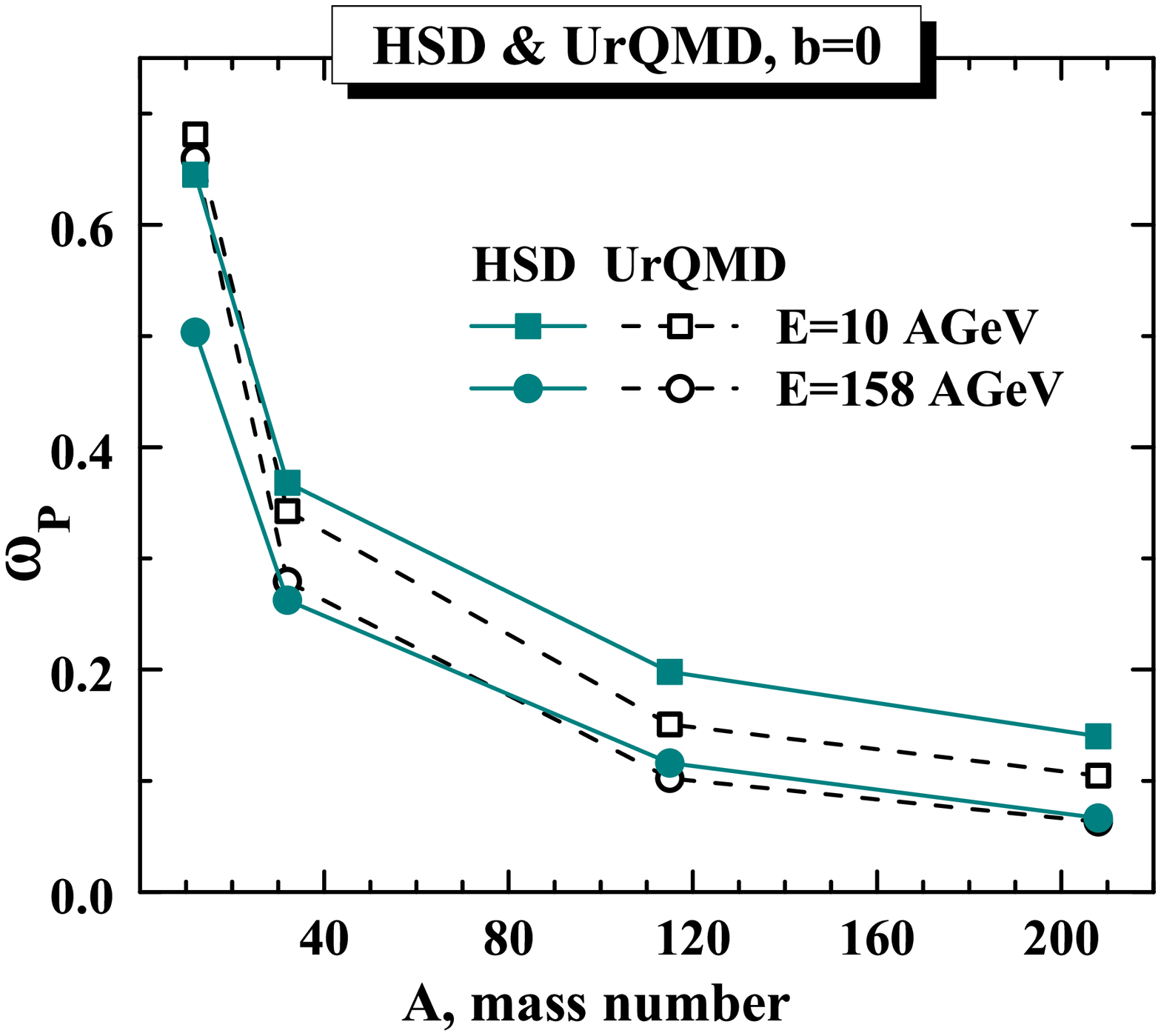,height=7cm}
\caption{(Color online) {\it Left:} Mean $\langle N_P\rangle$, divided by the
maximum number of participants $2A$ in  events with $b=0$ for
different nuclei at collision energies $E_{lab}$=10 and 158~AGeV.
{\it Right:} The scaled variance $\omega_P$ in events with $b=0$
for different nuclei at collision energies $E_{lab}$=10 and
158~AGeV. } \label{Np_distr}
\end{figure}

Let's start with the $b=0$ centrality selection criterium.  We
recall that the charged multiplicity fluctuations are closely
related to the fluctuations of the number of participants
\cite{KGB1,KGB3}. Therefore, it is useful to estimate the average
number of participants, $\langle N_P\rangle$, and the scaled
variances of its fluctuations, $\omega_P$, in $A+A$ collision
events  which satisfy the $b=0$ condition. The left panel in
Fig.~\ref{Np_distr} shows  the ratio, $\langle N_P\rangle/2$A, in
$A+A$ collisions with $b=0$ for different nuclei at collision
energies $E_{lab}=10$ and 158~AGeV. Both transport models (HSD and
UrQMD) show a monotonous increase of $\langle N_P\rangle/2$A with
collision energy for all nuclei in the energy range $10\div
158$~AGeV (Fig.~\ref{Np_distr}, {\it left}).  Correspondingly, the
fluctuations of the number of participants $\omega_P$ for all
nuclei become smaller with increasing collision energy
(Fig.~\ref{Np_distr}, {\it right}.).  As seen from
Fig.~\ref{Np_distr} ({\it left}) about 90\% of nucleons are
participants for Pb+Pb collisions with $b=0$. This number becomes
essentially smaller, about 60-70\%, for C+C collisions. One can
therefore expect that participant number fluctuations at $b=0$ are
small for heavy nuclei but strongly increase for light systems.
This is demonstrated in  Fig.~\ref{Np_distr} ({\it right}.):
$\omega_P$ is about 0.1$\div 0.2$ in Pb+Pb and In+In but becomes
much larger, $0.5\div$0.7, in C+C collisions.

One can conclude that the condition $b=0$ corresponds to `most
central' $A+A$ collisions only for nuclei with large atomic mass  number
(In and Pb). In this case the average number of participants is
close to its maximum value and its fluctuations are rather small.
However, in the studies of event-by-event multiplicity
fluctuations in the collisions of light nuclei (C and S) the
criterium $b=0$ is far from selecting the `most central' $A+A$
collisions.

\subsection{HSD and UrQMD Results for the Multiplicity Fluctuations
for $b$=0}

\begin{figure}[t!]
\epsfig{file=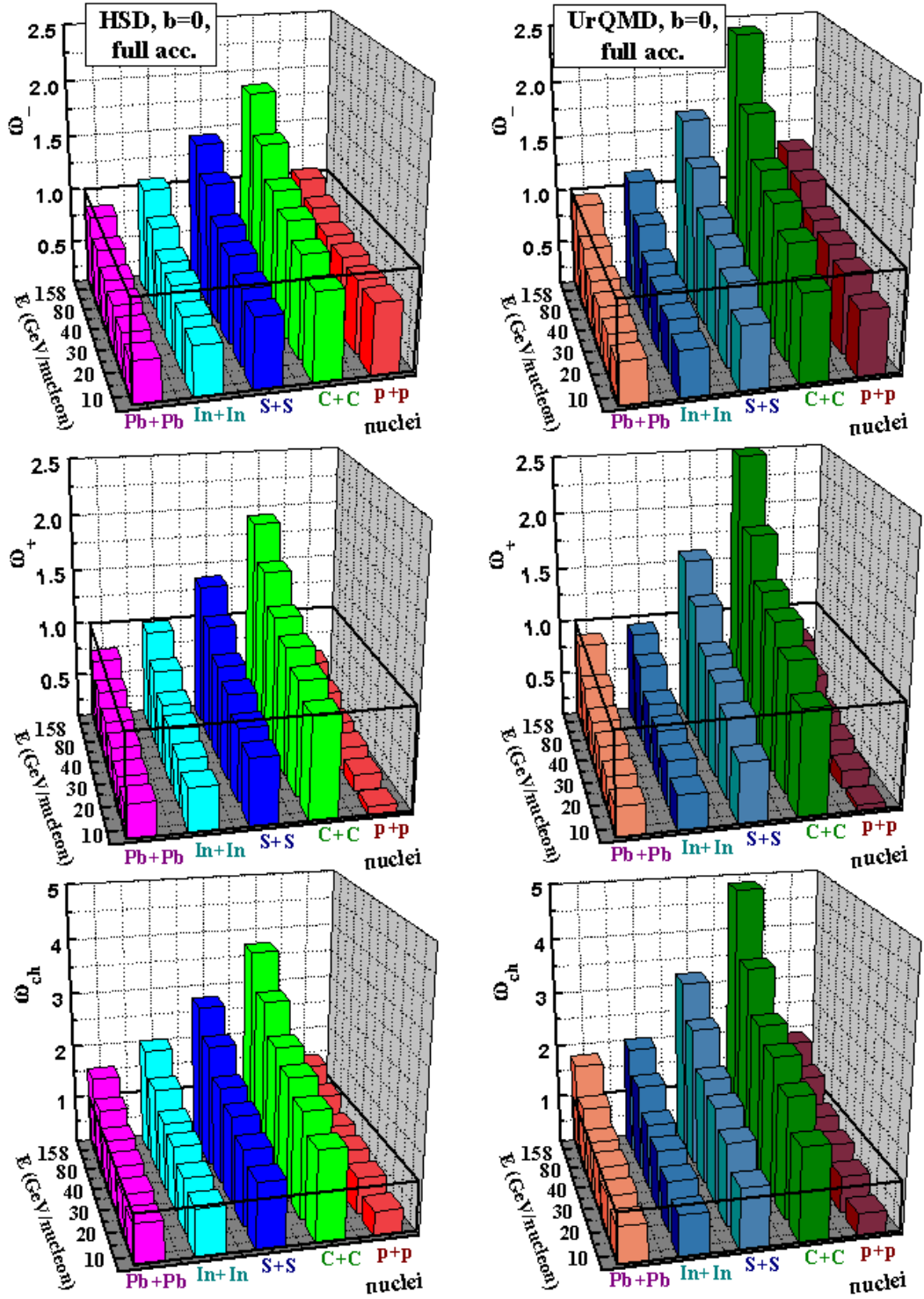,width=15cm} \vspace*{-5mm} \caption{(Color online) The
results of HSD ({\it left})  and UrQMD ({\it right}) simulations
for $\omega_-$ (top panel), $\omega_+$ (middle panel), and
$\omega_{ch}$ (lower panel) in p+p and central C+C, S+S, In+In,
Pb+Pb collisions at $E_{lab}=$~10, 20, 30, 40, 80, 158~AGeV. The
condition $b=0$ is used here as a criterium for centrality
selection.  There are no cuts in acceptance.} \label{wi_b0_f}
\end{figure}

\begin{figure}[ht!]
\epsfig{file=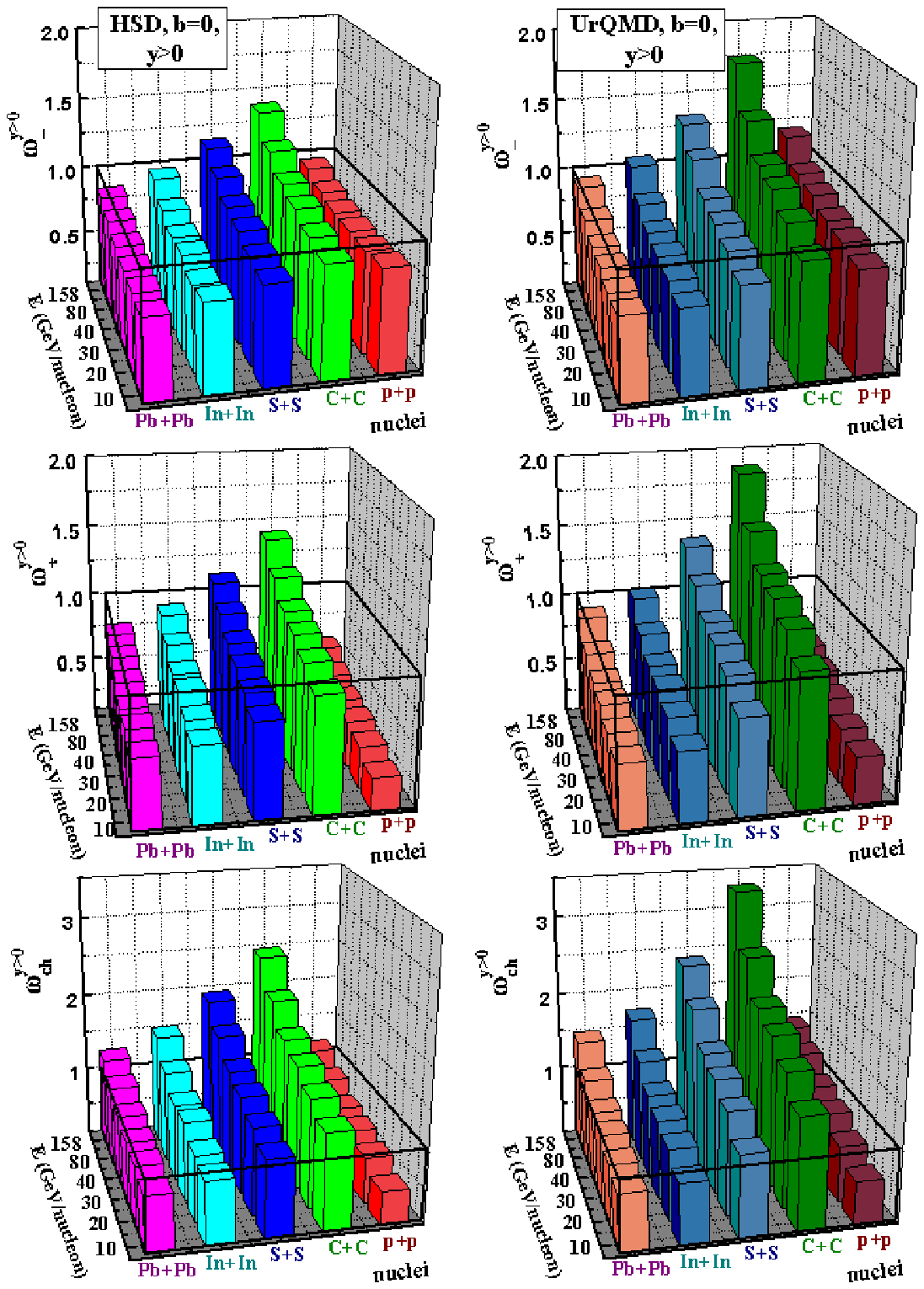,width=15cm} \caption{(Color online) The same as in
Fig.~\ref{wi_b0_f}, but only hadrons with positive c.m.
rapidities, $y>0$ (projectile hemisphere), are accepted. }
\label{wi_b0_y}
\end{figure}

Results of HSD and UrQMD transport model calculations for the
scaled variance of negative, $\omega_-$, positive, $\omega_+$, and
all charged, $\omega_{ch}$, hadrons are shown in
Figs.~\ref{wi_b0_f} and \ref{wi_b0_y} at different collision
energies, $E_{lab}=$~10, 20, 30, 40, 80, 158~AGeV, and for
different colliding nuclei,  C+C, S+S, In+In, Pb+Pb. The transport
model results correspond to collision events for zero impact
parameter, $b=0$. To make the picture more complete, the transport
model results for inelastic p+p collisions are shown too, for
reference.  Note that in our presentation - throughout the paper -
the proton spectators are not accounted for in the calculation of
$N_+$ and $N_{ch}$. Thus, proton spectators do not contribute to
$\omega_+$ and $\omega_{ch}$.

One sees a monotonic dependence of the multiplicity fluctuations
on both $E_{lab}$ and $A$: the scaled variances $\omega_-$, $\omega_+$,
and $\omega_{ch}$ increase with $E_{lab}$ and decrease with $A$. The
results for p+p collisions are different from those for light
ions. We note that within HSD and UrQMD a detailed comparison of the
multiplicity fluctuations in nucleon-nucleon inelastic collisions and
$b=0$ heavy-ion collisions (Pb+Pb and Au+Au), including the energy
dependence up to $\sqrt{s_{NN}}=200$~GeV, has been presented in
Refs.~\cite{KGB3,LB}.

Fig.~\ref{wi_b0_f} corresponds to the full $4\pi$ acceptance, i.e.
all particles are accepted without any cuts in phase space. In
actual experiments the detectors accept charged hadrons in limited
regions of  momentum space. Fig.~\ref{wi_b0_y} shows the HSD and
UrQMD results for multiplicity fluctuations in the projectile
hemisphere (i.e. positive rapidities, $y>0$ in the c.m. frame).
This corresponds to the maximal possible acceptance, up to 50\% of
all charged particles, by the optimized detectors of the NA61
Collaboration \cite{NA61}. One observes from Fig.~\ref{wi_b0_y}
that the energy and system size dependencies of the multiplicity
fluctuations in the projectile hemisphere $(y>0)$ become less
pronounced than in full $4\pi$ acceptance. Note also that the
centrality selection criterium $b=0$ keeps the symmetry between
the projectile and target hemispheres. Thus, the results for a
$y<0$ acceptance are identical to those for $y>0$ presented in
Fig.~\ref{wi_b0_y}.

\subsection{Comparison to the Independent Source Model}

The multiplicity fluctuations in elementary nucleon-nucleon
collisions and fluctuations of the number of nucleon participants
are presented in the right panels of Figs.~\ref{flucP} and
\ref{Np_distr}, respectively. Their combination explains the main
features of hadron multiplicity fluctuations in $A+A$ collisions
shown in Figs.~\ref{wi_b0_f} and \ref{wi_b0_y}, in particular, the
dependence on collision energy and atomic mass  number. They also
are responsible for the larger values of $\omega_i$ in the UrQMD
simulations in  comparison to those from HSD. To illustrate
this let us consider the model of independent sources (ISM).

The multiplicity fluctuations in $A+A$ collisions can be then written
according to the ISM as
(see e.g., Refs.~ \cite{rev1,KGB1,KGB3,KGB4,MGMG}),
\eq{
 \omega_i~=~\omega^*_i~+~n_i~ \omega_P~,
\label{WMod}}
where $\omega^*_i$ denotes  the fluctuations of the hadron multiplicity
from one source and the term $n_i~\omega_P$ gives additional
fluctuations due to the fluctuations of the number of sources. One
usually assumes that the number of sources is proportional to the
number of nucleon participants. The value of $n_i$ in Eq.~(\ref{WMod})
then is the average number of $i$'th particles per participant,
$n_i=\langle N_i\rangle /\langle N_P\rangle$, and $\omega_P$ equals the
scaled variance for the number of nucleon participants. Nucleon-nucleon
collisions, which are the weighted combinations of p+p, p+n, and
n+n reactions,   define the fluctuations $\omega_i^*$ from a single
source (see details in Refs.~\cite{KGB1,KGB3}).

\begin{figure}[ht!]
\epsfig{file=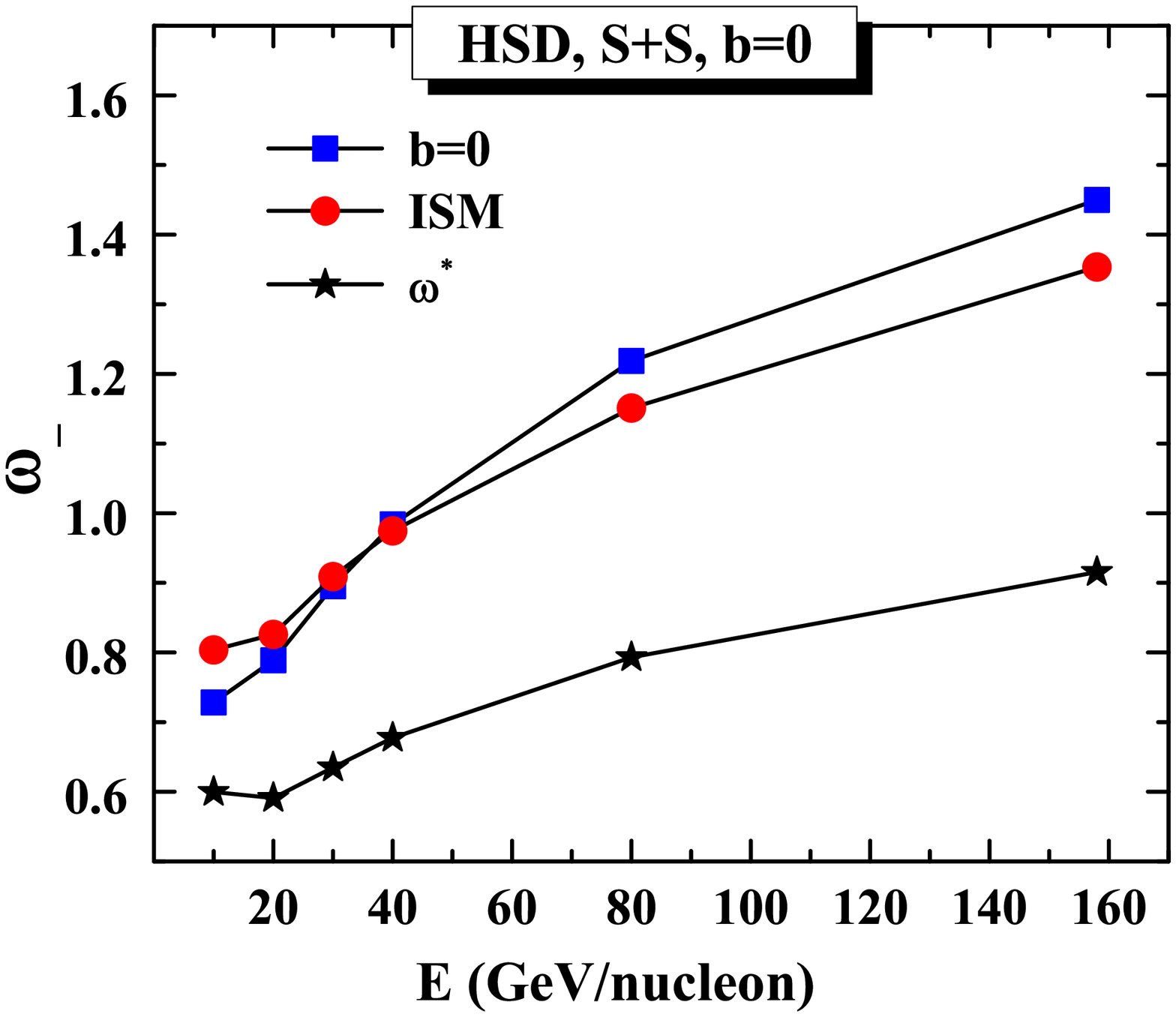,height=7cm}
\epsfig{file=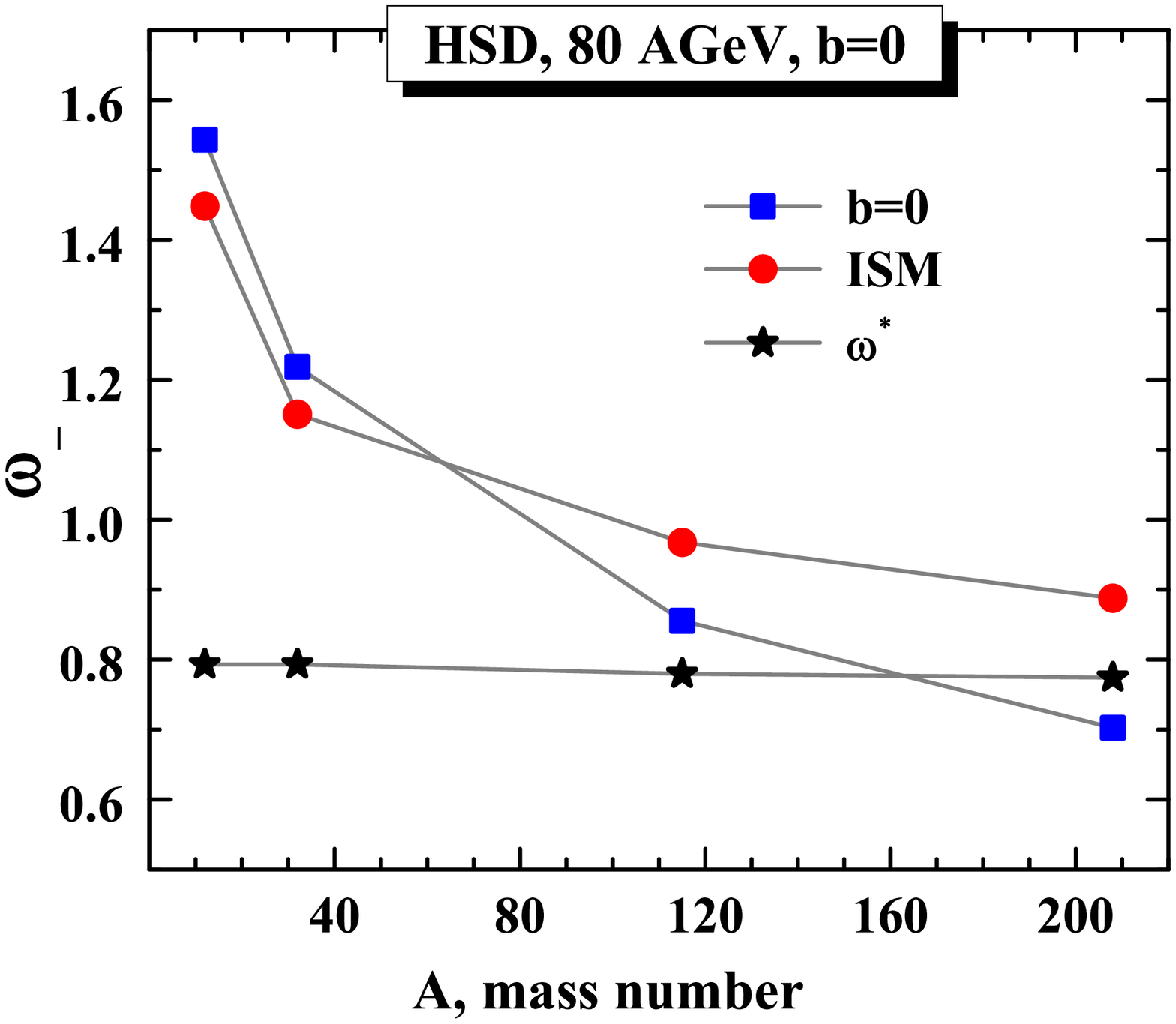,height=7cm} \caption{(Color online)  The {\it left} panel
illustrates the energy dependence of $\omega_-$ in S+S  collisions
at $b=0$ in the full $4\pi$ acceptance, the {\it right} panel --
the $\omega_-$ dependence on atomic mass  number at $E_{lab}$=80~AGeV.
The HSD results are shown by the squares while the circles
correspond to Eq.~(\ref{WMod}) of the model of independent
sources. The stars show the first term, $\omega^*_-$, in the {\it
right}. of Eq.~(\protect\ref{WMod}) -- scaled variance for
negative hadrons in nucleon-nucleon collisions, where the values
of $\omega_-^*$, $n_i$, and $\omega_P$ are calculated within HSD.
} \label{MoIS}
\end{figure}

In Fig.~\ref{MoIS} the HSD results for $\omega_i$ in $A+A$
collisions at $b=0$ are compared to the ISM -- Eq.~(\ref{WMod}). One
concludes that the transport model results for the multiplicity
fluctuations are in qualitative agreement with Eq.~(\ref{WMod}) of
the independent source model. Both $n_i$ and $\omega_i^*$ increase
strongly with collision energy as seen from Fig.~\ref{flucP}. This
explains, due to Eq.~(\ref{WMod}), the monotonous increase with
energy of the scaled variances $\omega_i$ in $A+A$ collisions at
$b=0$ seen in Figs.~\ref{wi_b0_f} and \ref{wi_b0_y}. Note that
$\omega_P$ at $b=0$ decreases with collision energy as shown in
Fig.~\ref{Np_distr}, {\it right}. This, however, does not
compensate a strong increase of both $n_i$ and $\omega_i^*$. The
atomic mass  number dependence of the scaled variances $\omega_i$ in
$A+A$ collisions with $b=0$ follows from the A-dependence of
$\omega_P$. Fig.~\ref{Np_distr} ({\it right}) demonstrates a
strong increase of $\omega_P$ for light nuclei. This, due to
Eq.~(\ref{WMod}), is transformed to the corresponding behavior of
$\omega_i$ seen in Figs.~\ref{wi_b0_f} and \ref{wi_b0_y}.

\section{Multiplicity fluctuations in 1\%  most central collisions}

We consider now the centrality selection procedure by fixing the
number of projectile participants $N_P^{proj}$. This corresponds
to the real situation of $A+A$ collisions in fixed target
experiments. As a first step we simulate in HSD and UrQMD the
minimal bias events - which correspond to an all impact parameter
sample - and calculate the event distribution over the number of
participants $N_{part}$. Then, we select  1\% most central
collisions which correspond to the largest values of $N_P^{proj}$.
In such a sample of A+A collisions events with largest
$N_P^{proj}$ from different impact parameters can contribute.
After that we calculate the values of $\omega_P$ in these samples.
Note  that even for a fixed number of $N_P^{proj}$ the number of
target participants $N_P^{targ}$ fluctuates. Thus, the total
number of participants, $N_P=N_P^{targ}+N_P^{proj}$, fluctuates
too. In our 1\% sample, both $N_P^{targ}$ and $N_P^{proj}$
fluctuate. Besides, there are correlations between $N_P^{targ}$
and $N_P^{proj}$.

\begin{figure}[ht!]
\epsfig{file=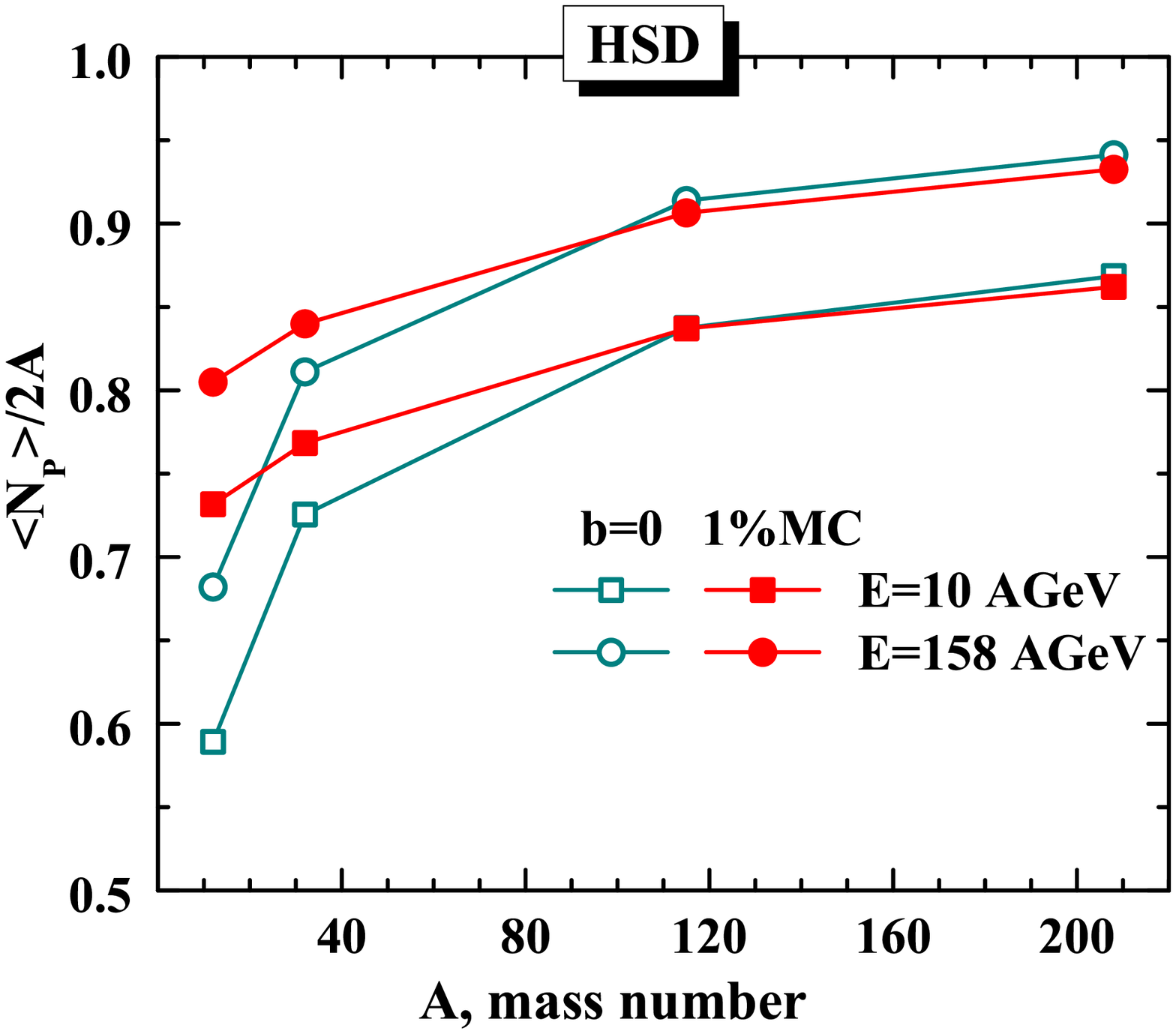,height=7cm} \epsfig{file=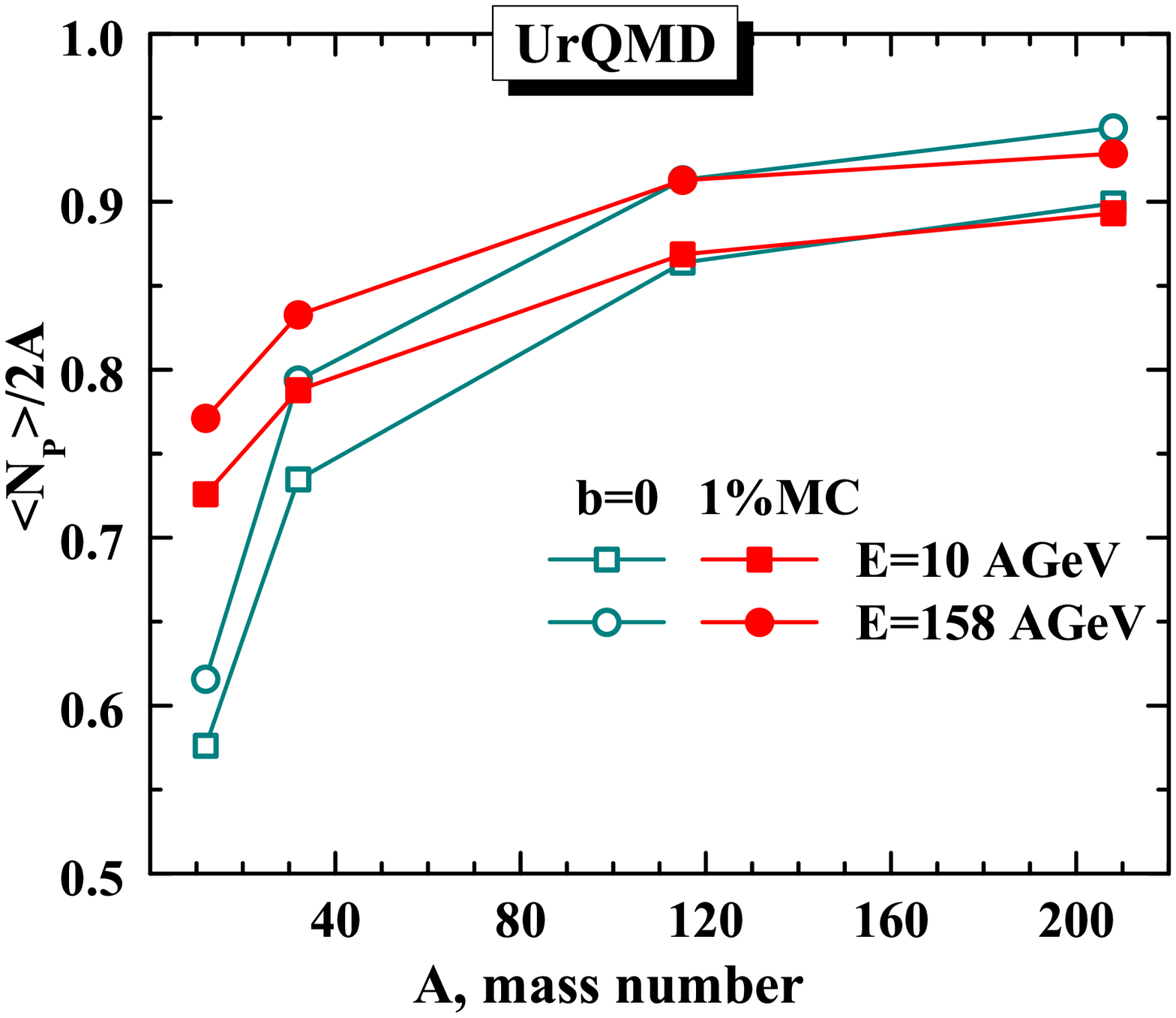,height=7cm}
\epsfig{file=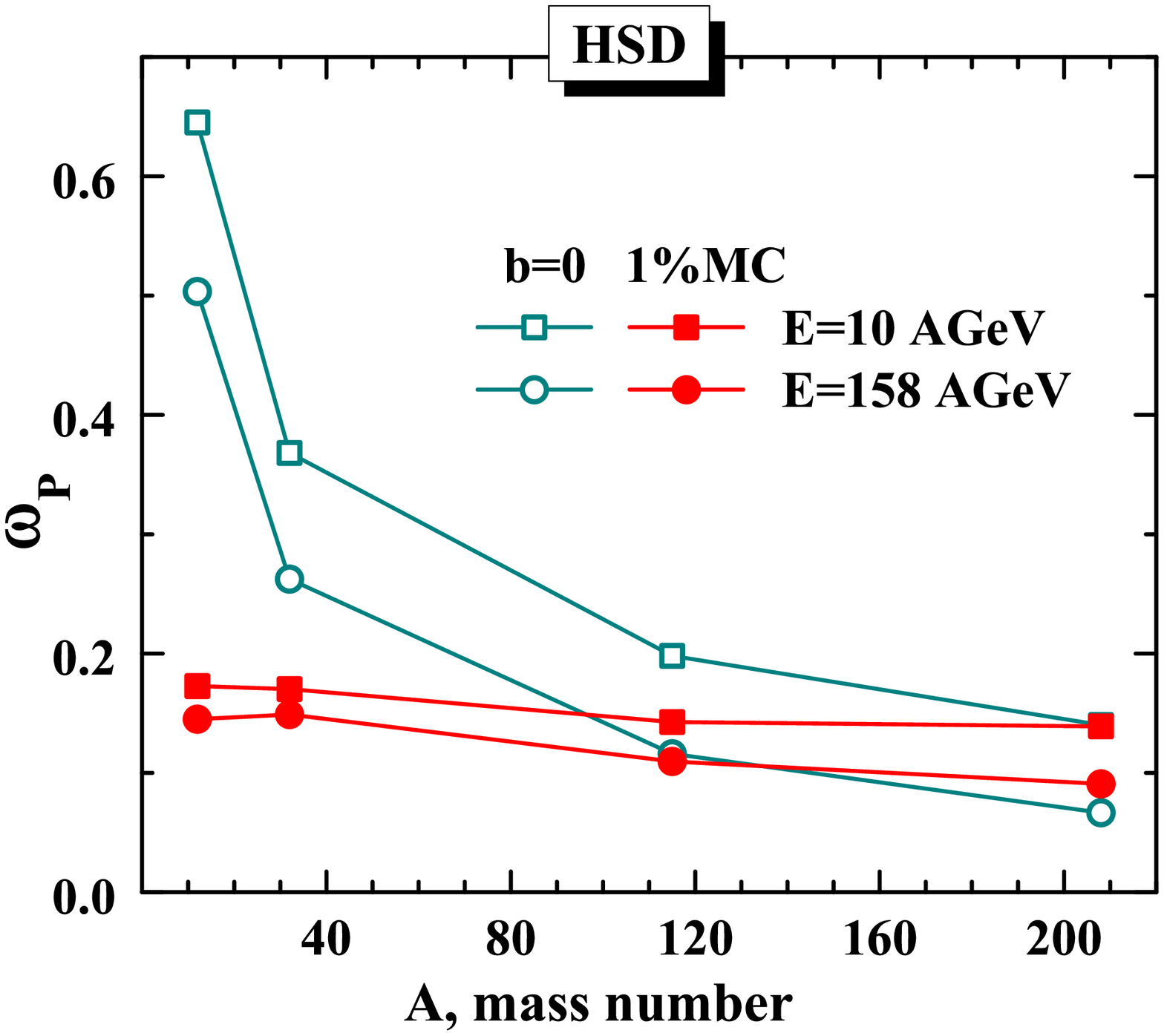,height=7cm} \epsfig{file=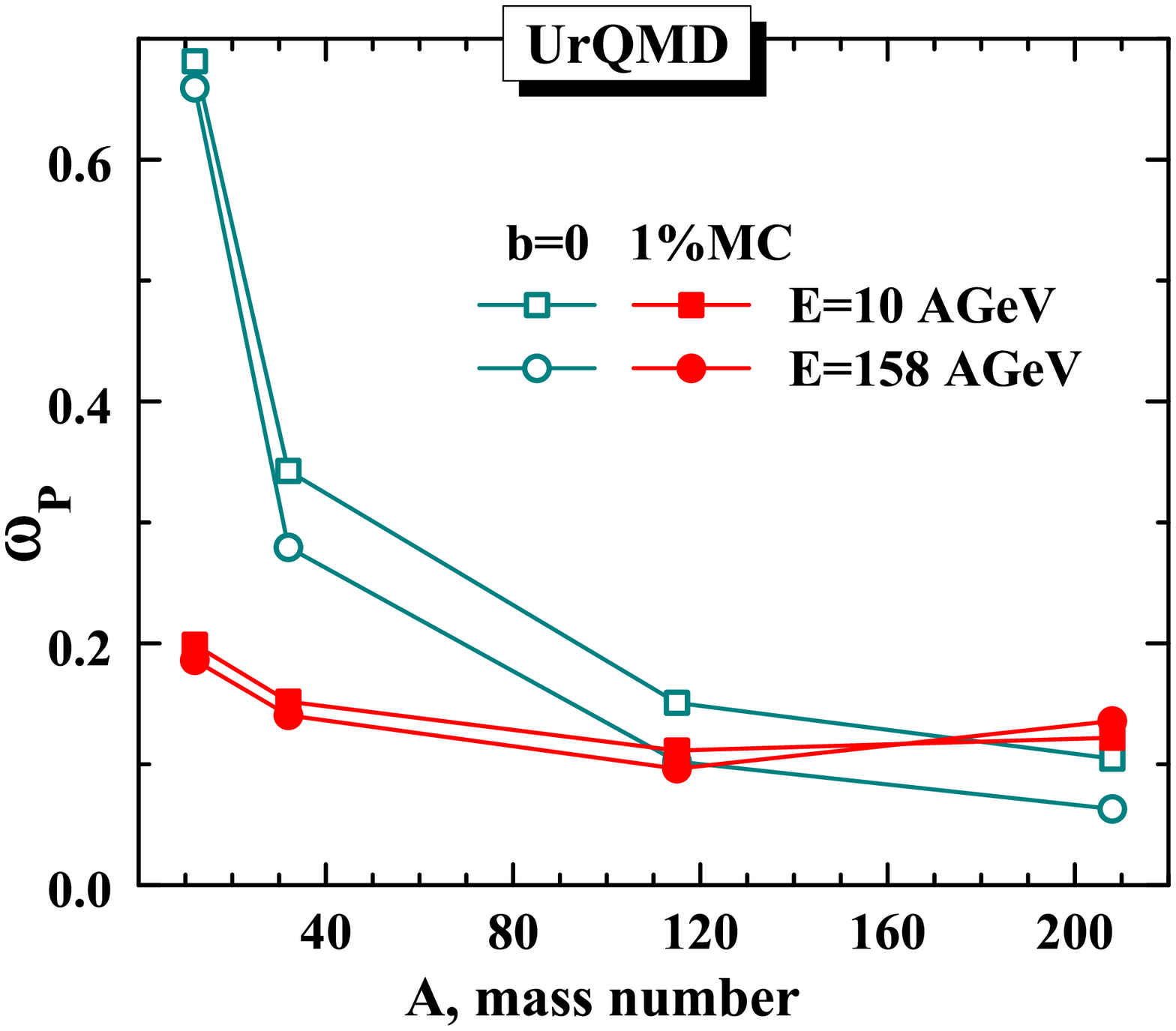,height=7cm}
\caption{(Color online)  The HSD ({\it left}) and UrQMD ({\it right}) results for
the ratio $\langle N_P\rangle/2A$ (the upper panel) and the scaled
variance of the participant number fluctuations, $\omega_P$ (the
lower panel), for the 1\% most central collisions selected by the
largest values of $N_P^{proj}$ (full symbols), for different
nuclei at collision energies $E_{lab}$=10 and 158~AGeV. The open
symbols present the results of Fig.~\ref{Np_distr} ({\it right})
for $b=0$. } \label{Np_distr2}
\end{figure}

Fig.~\ref{Np_distr2} shows the ratio $\langle N_P\rangle/2A$ and
the scaled variance, $\omega_P$, for 1\% most central collisions
selected by the largest values of $N_P^{proj}$. These results are
compared with those for the $b=0$ centrality selection. For heavy
nuclei, like In and Pb, one finds no essential differences between
these two criteria of centrality selection. However, the 1\%
centrality trigger defined by the largest values of $N_P^{proj}$
looks much more rigid for light ions (S and C).  In this case, the
ratio $\langle N_P\rangle/2A$ is larger, and $\omega_P$ is
essentially smaller than for the criterion  $b=0$. As a result, the 1\%
centrality trigger by the largest values of $N_P^{proj}$  leads to
a rather weak $A$-dependence of $\omega_P$.

Some comments are appropriate at this point. Let us define the
centrality $c(N)$ as a percentage of events with a multiplicity
larger than $N$ (this can be the number of produced hadrons,
number of participants, etc.).  It was argued in Ref.~\cite{BF}
that a selection of $c(N)$ of most central $A+A$ collisions is
equivalent to restricting the impact parameter, $b<b(N)$, with,
\eq{\label{centrality} b(N)~=~
\sqrt{\frac{\sigma_{inel}}{\pi}~c(N)}~, }
where $\sigma_{inel}$ is the total inelastic A+A cross section.
Thus, the centrality criterion by the multiplicity $N$ is
equivalent to the geometrical criterion by the impact parameter
$b$. Moreover, the result (\ref{centrality}) does not depend on
the specific observable $N$ used to define the $c$-percentage of
most central A+A collisions. Eq.~(\ref{centrality}) should remain
the same for any observable $N$ which is a monotonic function of
$b$. Therefore, the relation (\ref{centrality}) reduces any
centrality selection to the geometrical one. This result was
obtained in Ref.~\cite{BF} by neglecting the fluctuations of
multiplicity $N$ at a given value of $b$. This is valid if $c$ is
not too small and the colliding nuclei are not too light. In the
sample of A+A events with 1\% of largest $N_P^{proj}$, the
relation (\ref{centrality}) can not be applied for S+S and C+C
collisions. The average value of $\langle N_P^{proj}\rangle$ even
for $b=0$ is essentially smaller than its maximal value A.
To form the sample with 1\% largest $N_P^{proj}$, one needs at several
fixed values of $b$ to take into account the {\it fluctuations}
with $N_P^{proj} > \langle N_P^{proj} (b)\rangle$.

Just the
fluctuations of the number of nucleon participants form the 1\%
sample with largest $N_P^{proj}$ values.

\begin{figure}[!]
\epsfig{file=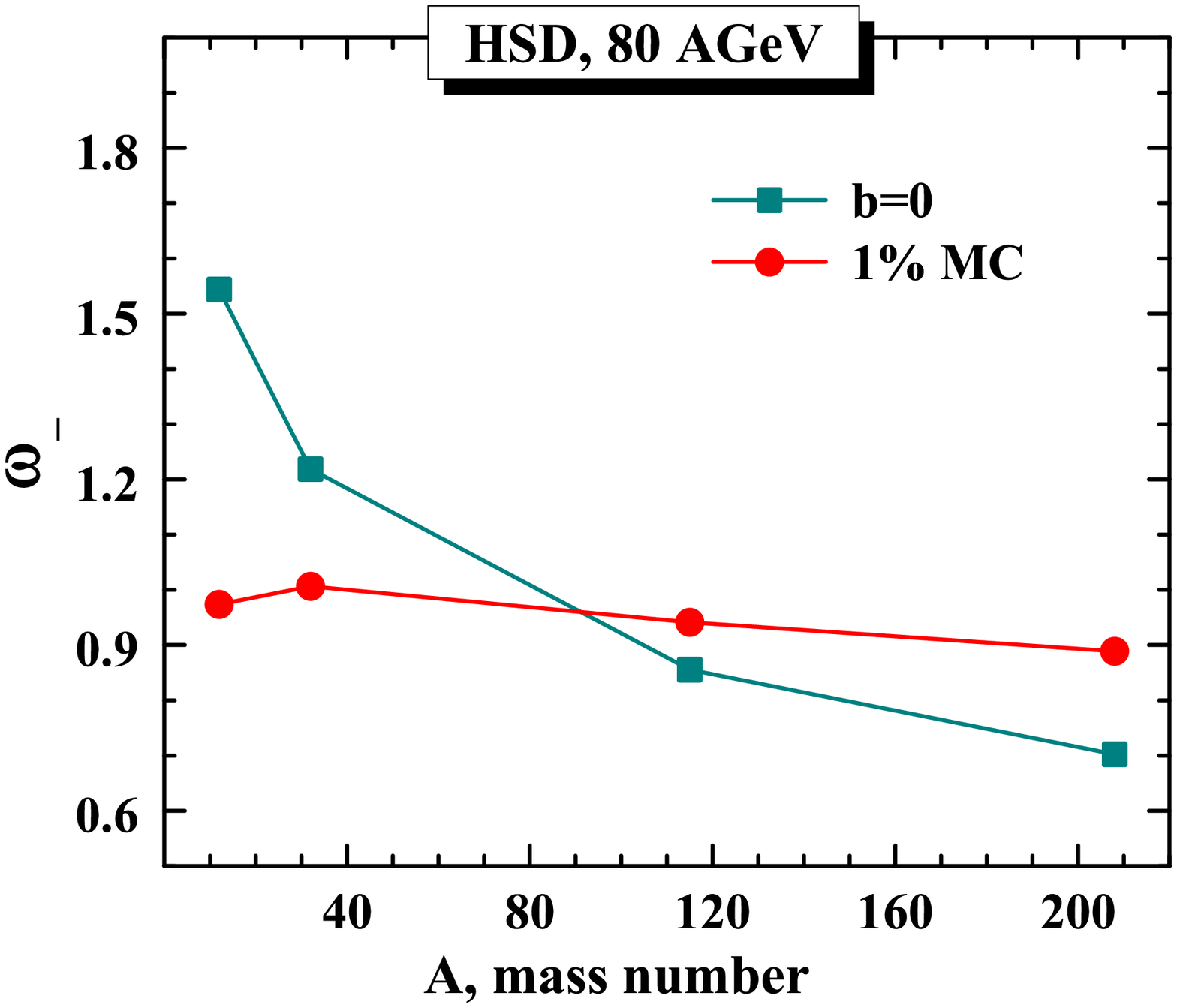,height=7cm}
\epsfig{file=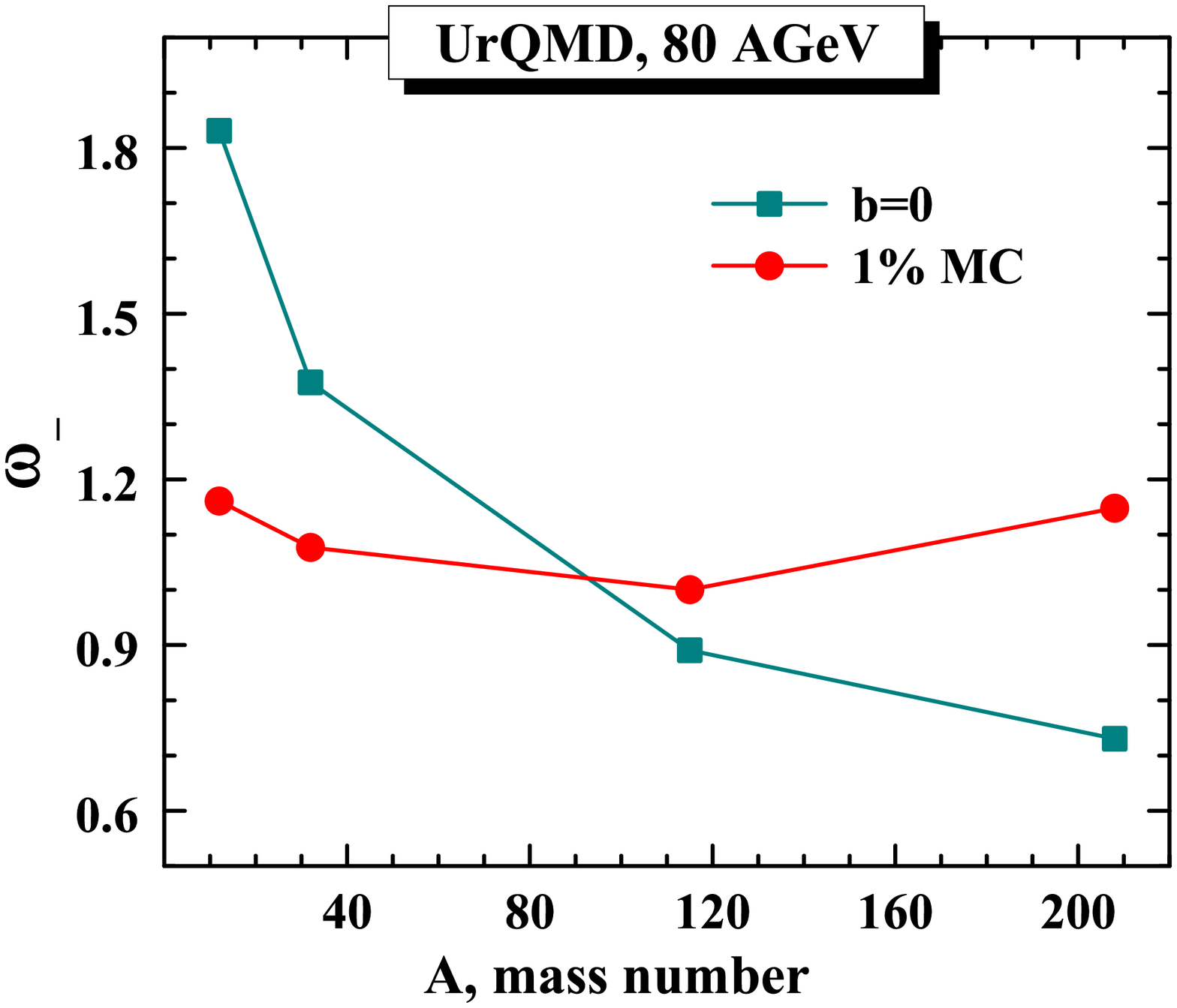,height=7cm} \caption{(Color online)  The  dependence
of $\omega_-$  on atomic mass  number at $E_{lab}$=80~AGeV for the HSD
({\it left}) and UrQMD ({\it left}) simulations. The squares
correspond to $b=0$, and circles to 1\% largest $N_P^{proj}$. }
\label{A-dep}
\end{figure}

Fig.~\ref{A-dep} shows a comparison of the A-dependence of
$\omega_-$ in the transport models for two different samples of
the collision events: for $b=0$ and for the 1\% of events with
largest $N_P^{proj}$ values.  One can see that the multiplicity
fluctuations are rather different in these two samples.  Moreover,
these  differences are  in the opposite directions for heavy
nuclei and for light nuclei. For light nuclei, $\omega_-$ is
essentially smaller in the 1\% sample with largest $N_P^{proj}$
values, whereas for heavy nuclei the smaller fluctuations
correspond to $b=0$ events. Note that in the 1\% sample with
largest $N_P^{proj}$ values the A-dependence of multiplicity
fluctuations becomes much weaker. In this case, a strong increase
of the multiplicity fluctuations for light nuclei, seen for $b=0$,
disappears.

\begin{figure}[!]
\epsfig{file=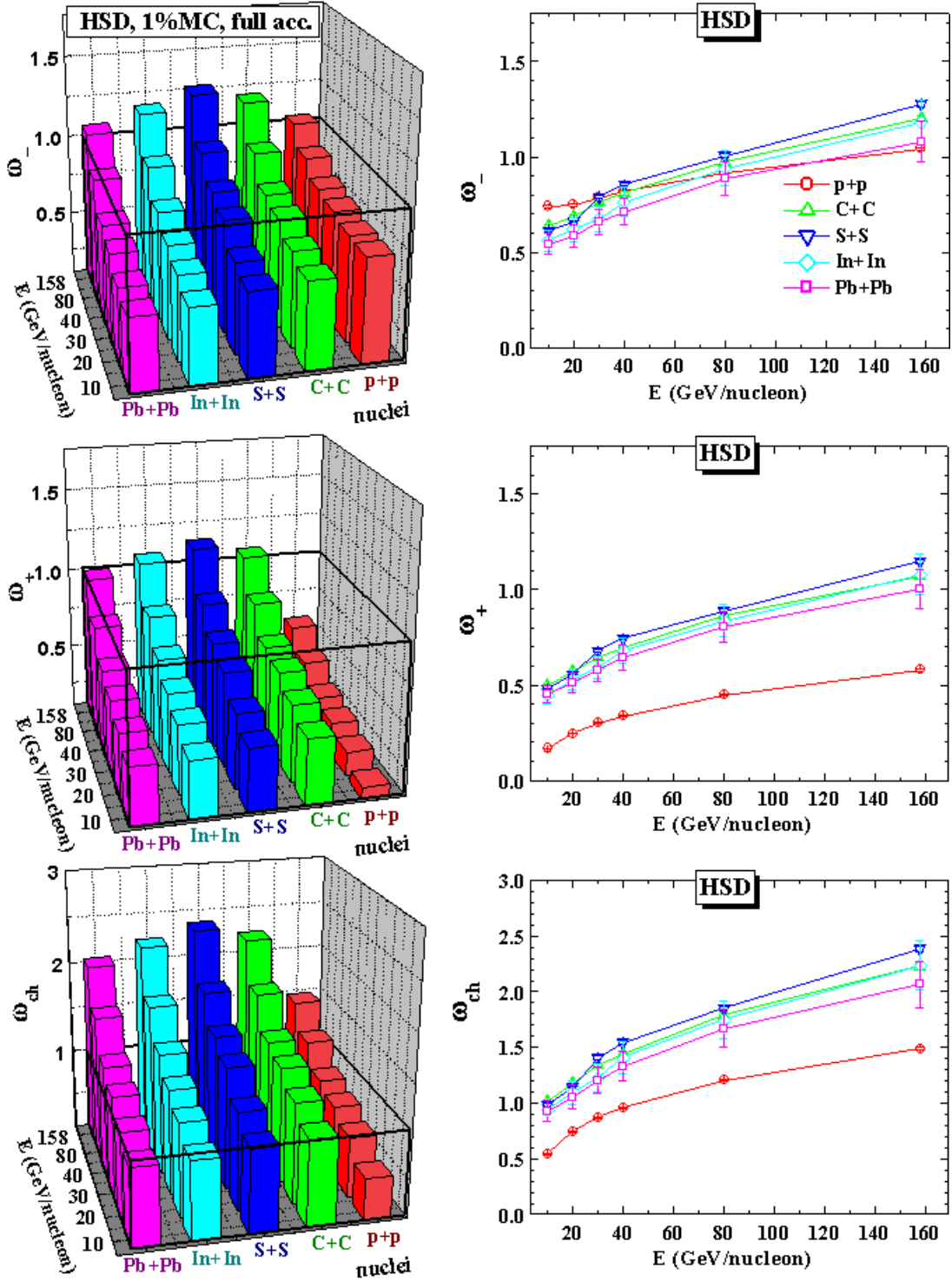,width=15cm} \caption{(Color online) The HSD results for
$\omega_-$ (upper panel), $\omega_+$ (middle panel), and
$\omega_{ch}$ (lower panel) in $A+A$ and p+p collisions for the
full $4\pi$ acceptance in 3D ({\it left}) and 2D ({\it right}.)
projection. The 1\%  most central C+C, S+S, In+In, and Pb+Pb
collisions are selected by choosing the largest values of
$N_P^{proj}$ at different collision energies $E_{lab}$=10, 20, 30,
40, 80, 158~AGeV.  The errorbars indicate the estimated
uncertainties in the model calculations. The HSD results from
inelastic p+p collisions are the same as in Fig.~\ref{wi_b0_f}. }
\label{wi_1pc_f}
\end{figure}

\begin{figure}[!]
\epsfig{file=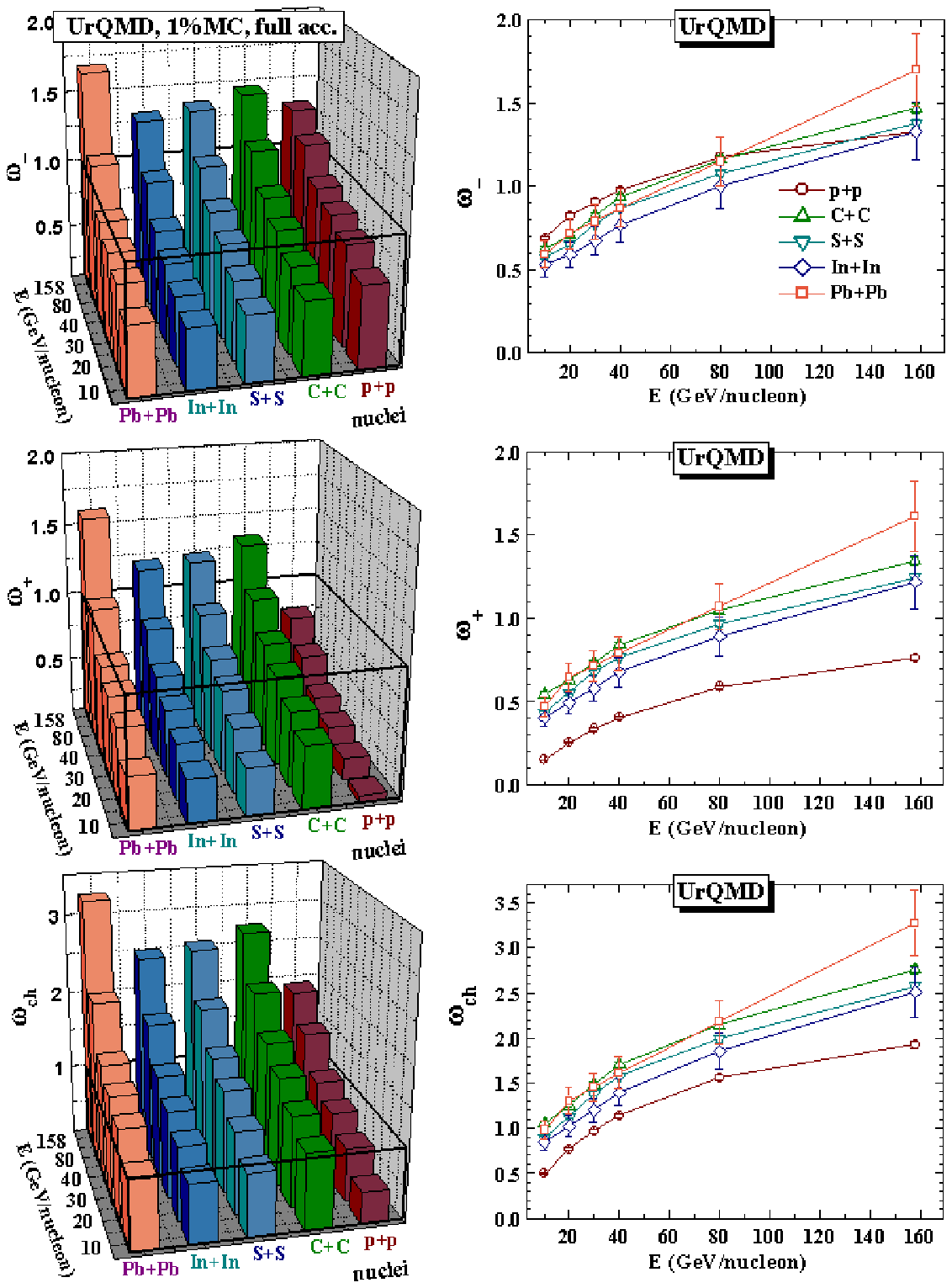,width=16.2cm} \caption{(Color online) The same as in
Fig.~\ref{wi_1pc_f}, but for the UrQMD. } \label{wi_1pc_f2}
\end{figure}

\begin{figure}[!]
\epsfig{file=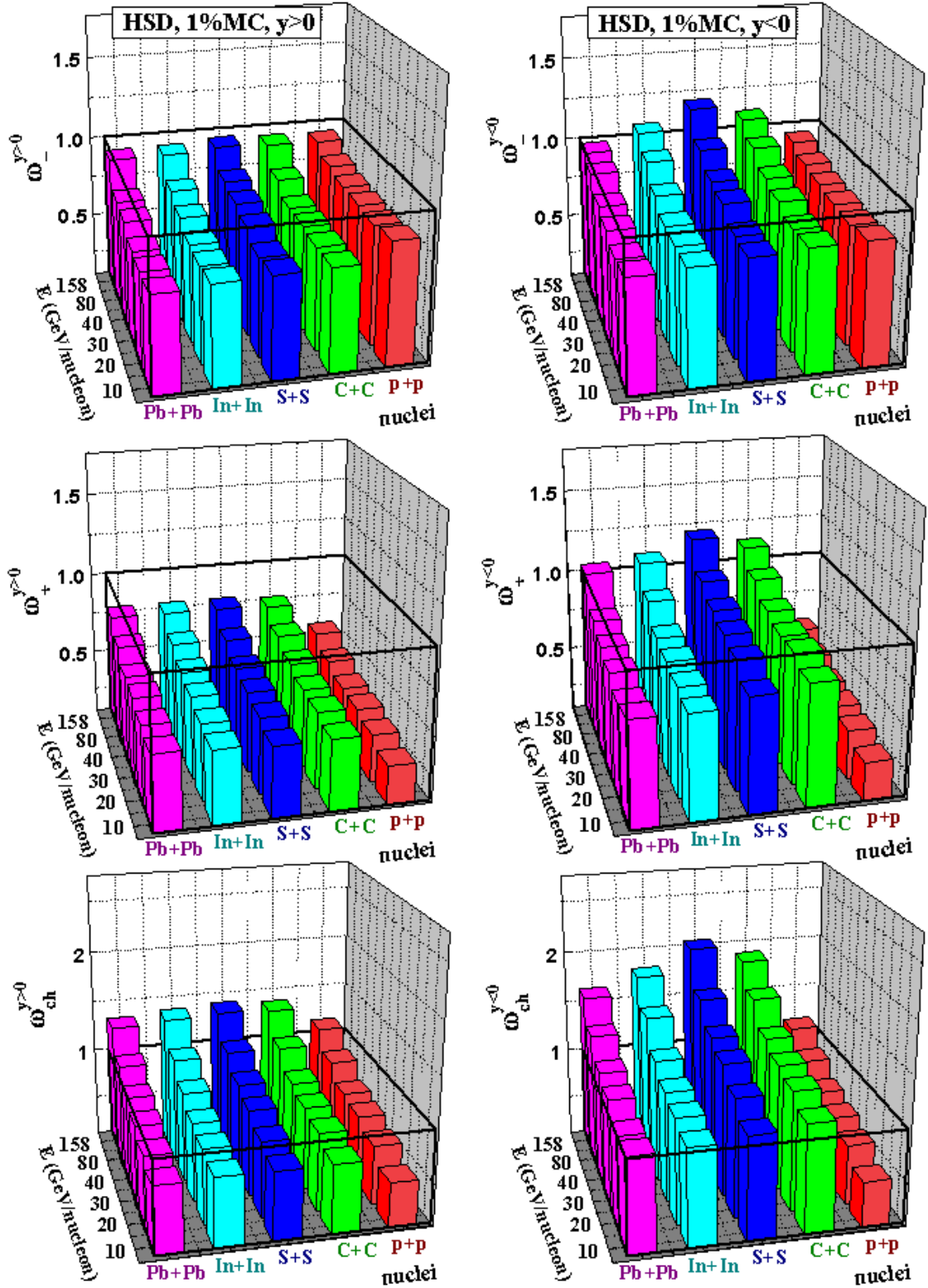,width=15.8cm} \caption{(Color online) The same as in
Fig.~\ref{wi_1pc_f}, but for  final hadrons accepted in the
projectile hemisphere, $y>0$ ({\it left}), and in the target
hemisphere, $y<0$ ({\it right}). The HSD results in inelastic p+p
collisions are the same as in Fig.~\ref{wi_b0_y}. }
\label{wi_1pc_y}
\end{figure}

\begin{figure}[!]
\epsfig{file=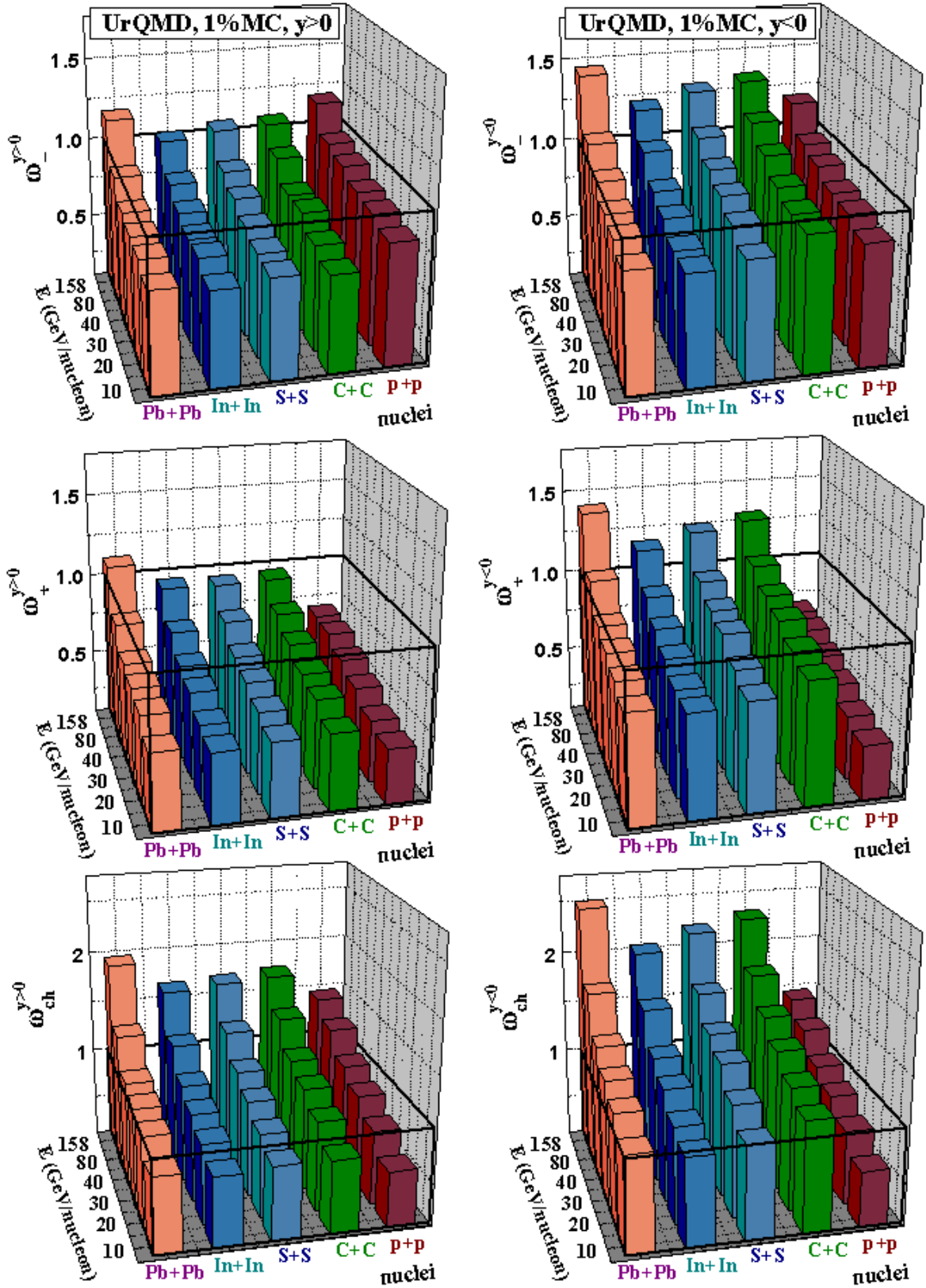,width=16.5cm} \caption{(Color online) The same as in
Fig.~\ref{wi_1pc_y}, but for the UrQMD. } \label{wi_1pc_y2}
\end{figure}

For the 1\% most central $A+A$ collision events - selected by the
largest values of $N_P^{proj}$ - the HSD multiplicity fluctuations
are shown in Figs.~\ref{wi_1pc_f} and \ref{wi_1pc_y} and the
corresponding UrQMD results are shown in Figs.~\ref{wi_1pc_f2} and
\ref{wi_1pc_y2}. The results from both models are also presented
in Table~\ref{table1}. The model uncertainties are shown as
errorbars in Figs.~\ref{wi_1pc_f} and \ref{wi_1pc_f2}.  For light
nuclei (S and C) the multiplicity fluctuations in the samples of
1\%  most central collisions are smaller than in the $b=0$
selection and the atomic mass  number dependencies become less
pronounced (compare Figs.~\ref{wi_1pc_f} and~\ref{wi_1pc_f2} with
Fig.~\ref{wi_b0_f}). This is because the participant number
fluctuations $\omega_P$ have now essentially smaller A-dependence,
as seen in Fig.~\ref{Np_distr2}.
\begin{table}[!]
\begin{center}
\begin{tabular}{||c|c||c||c||c||c||c||} \hline\hline
\multicolumn{2}{||c||}{} & p+p & C+C & S+S & In+In & Pb+Pb \\
\multicolumn{2}{||c||}{} &
 \begin{tabular}{p{2em}|p{2em}|p{2em}} \hline \hspace{0.5em} - & \hspace{0.4em} + & \hspace{0.4em} ch \\\end{tabular} &
 \begin{tabular}{p{2em}|p{2em}|p{2em}} \hline \hspace{0.5em} - & \hspace{0.4em} + & \hspace{0.4em} ch \\\end{tabular} &
 \begin{tabular}{p{2em}|p{2em}|p{2em}} \hline \hspace{0.5em} - & \hspace{0.4em} + & \hspace{0.4em} ch \\\end{tabular} &
 \begin{tabular}{p{2em}|p{2em}|p{2em}} \hline \hspace{0.5em} - & \hspace{0.4em} + & \hspace{0.4em} ch \\\end{tabular} &
 \begin{tabular}{p{2em}|p{2em}|p{2em}} \hline \hspace{0.5em} - & \hspace{0.4em} + & \hspace{0.4em} ch \\\end{tabular} \\
\hline \hline 
\raisebox{-3em}{\rotatebox{90}{HSD: full acc.}}
&\begin{tabular}{r}10\\\hline 20\\\hline 30\\\hline 40\\\hline 80\\\hline 158\\\end{tabular}&
                  \begin{tabular}{p{2em}|p{2em}|p{2em}} 
    \ 0.74 &\ 0.17 &\ 0.54 \\\hline
    \ 0.75 &\ 0.25 &\ 0.74 \\\hline
    \ 0.79 &\ 0.30 &\ 0.87 \\\hline
    \ 0.82 &\ 0.34 &\ 0.96 \\\hline
    \ 0.92 &\ 0.45 &\ 1.21 \\\hline
    \ 1.04 &\ 0.58 &\ 1.49 \\
  \end{tabular} & \begin{tabular}{p{2em}|p{2em}|p{2em}} 
    \ 0.64 &\ 0.50 &\ 1.02 \\\hline
    \ 0.69 &\ 0.57 &\ 1.18 \\\hline
    \ 0.76 &\ 0.64 &\ 1.34 \\\hline
    \ 0.81 &\ 0.69 &\ 1.44 \\\hline
    \ 0.97 &\ 0.86 &\ 1.79 \\\hline
    \ 1.20 &\ 1.07 &\ 2.23 \\
  \end{tabular} & \begin{tabular}{p{2em}|p{2em}|p{2em}} 
    \ 0.61 &\ 0.48 &\ 0.99 \\\hline
    \ 0.66 &\ 0.55 &\ 1.14 \\\hline
    \ 0.79 &\ 0.68 &\ 1.41 \\\hline
    \ 0.86 &\ 0.75 &\ 1.54 \\\hline
    \ 1.01 &\ 0.89 &\ 1.85 \\\hline
    \ 1.28 &\ 1.15 &\ 2.39 \\
  \end{tabular} & \begin{tabular}{p{2em}|p{2em}|p{2em}} 
    \ 0.56 &\ 0.45 &\ 0.94 \\\hline
    \ 0.62 &\ 0.53 &\ 1.10 \\\hline
    \ 0.67 &\ 0.59 &\ 1.22 \\\hline
    \ 0.76 &\ 0.68 &\ 1.41 \\\hline
    \ 0.94 &\ 0.84 &\ 1.75 \\\hline
    \ 1.18 &\ 1.08 &\ 2.24 \\
  \end{tabular} & \begin{tabular}{p{2em}|p{2em}|p{2em}} 
    \ 0.54 &\ 0.45 &\ 0.93 \\\hline
    \ 0.58 &\ 0.51 &\ 1.05 \\\hline
    \ 0.66 &\ 0.58 &\ 1.20 \\\hline
    \ 0.71 &\ 0.64 &\ 1.33 \\\hline
    \ 0.89 &\ 0.80 &\ 1.67 \\\hline
    \ 1.08 &\ 1.00 &\ 2.07 \\
  \end{tabular} \\
\hline \hline 
\raisebox{-2em}{\rotatebox{90}{HSD: $y>0$}}
&\begin{tabular}{r} 10 \\\hline 20\\\hline 30\\\hline 40 \\\hline 80 \\\hline 158 \\\end{tabular}&
                  \begin{tabular}{p{2em}|p{2em}|p{2em}} 
    \ 0.85 &\ 0.33 &\ 0.53 \\\hline
    \ 0.83 &\ 0.39 &\ 0.68 \\\hline
    \ 0.83 &\ 0.42 &\ 0.77 \\\hline
    \ 0.84 &\ 0.44 &\ 0.82 \\\hline
    \ 0.88 &\ 0.49 &\ 0.96 \\\hline
    \ 0.94 &\ 0.58 &\ 1.16 \\
  \end{tabular} & \begin{tabular}{p{2em}|p{2em}|p{2em}} 
    \ 0.73 &\ 0.53 &\ 0.77 \\\hline
    \ 0.73 &\ 0.56 &\ 0.86 \\\hline
    \ 0.74 &\ 0.58 &\ 0.91 \\\hline
    \ 0.74 &\ 0.62 &\ 0.97 \\\hline
    \ 0.82 &\ 0.69 &\ 1.16 \\\hline
    \ 0.94 &\ 0.78 &\ 1.39 \\
  \end{tabular} & \begin{tabular}{p{2em}|p{2em}|p{2em}} 
    \ 0.72 &\ 0.52 &\ 0.76 \\\hline
    \ 0.70 &\ 0.56 &\ 0.83 \\\hline
    \ 0.76 &\ 0.59 &\ 0.94 \\\hline
    \ 0.80 &\ 0.62 &\ 1.01 \\\hline
    \ 0.83 &\ 0.70 &\ 1.17 \\\hline
    \ 0.96 &\ 0.79 &\ 1.41 \\
  \end{tabular} & \begin{tabular}{p{2em}|p{2em}|p{2em}} 
    \ 0.71 &\ 0.54 &\ 0.77 \\\hline
    \ 0.68 &\ 0.56 &\ 0.82 \\\hline
    \ 0.69 &\ 0.60 &\ 0.87 \\\hline
    \ 0.74 &\ 0.63 &\ 0.98 \\\hline
    \ 0.80 &\ 0.70 &\ 1.13 \\\hline
    \ 0.93 &\ 0.80 &\ 1.38 \\
  \end{tabular} & \begin{tabular}{p{2em}|p{2em}|p{2em}} 
    \ 0.68 &\ 0.56 &\ 0.76 \\\hline
    \ 0.69 &\ 0.59 &\ 0.82 \\\hline
    \ 0.69 &\ 0.63 &\ 0.90 \\\hline
    \ 0.75 &\ 0.64 &\ 0.98 \\\hline
    \ 0.77 &\ 0.71 &\ 1.11 \\\hline
    \ 0.93 &\ 0.80 &\ 1.34 \\
  \end{tabular} \\
\hline \hline 
\raisebox{-3em}{\rotatebox{90}{UrQMD: full acc.}}
&\begin{tabular}{r} 10 \\\hline 20\\\hline 30\\\hline 40 \\\hline 80 \\\hline 158 \\\end{tabular}&
                  \begin{tabular}{p{2em}|p{2em}|p{2em}} 
    \ 0.69 &\ 0.15 &\ 0.49 \\\hline
    \ 0.82 &\ 0.25 &\ 0.77 \\\hline
    \ 0.90 &\ 0.34 &\ 0.96 \\\hline
    \ 0.97 &\ 0.41 &\ 1.13 \\\hline
    \ 1.18 &\ 0.59 &\ 1.56 \\\hline
    \ 1.33 &\ 0.76 &\ 1.93 \\
  \end{tabular} & \begin{tabular}{p{2em}|p{2em}|p{2em}} 
    \ 0.62 &\ 0.54 &\ 1.05 \\\hline
    \ 0.71 &\ 0.63 &\ 1.25 \\\hline
    \ 0.83 &\ 0.73 &\ 1.48 \\\hline
    \ 0.93 &\ 0.84 &\ 1.70 \\\hline
    \ 1.16 &\ 1.05 &\ 2.15 \\\hline
    \ 1.47 &\ 1.34 &\ 2.76 \\
  \end{tabular} & \begin{tabular}{p{2em}|p{2em}|p{2em}} 
    \ 0.57 &\ 0.43 &\ 0.89 \\\hline
    \ 0.65 &\ 0.55 &\ 1.12 \\\hline
    \ 0.77 &\ 0.68 &\ 1.38 \\\hline
    \ 0.87 &\ 0.76 &\ 1.57 \\\hline
    \ 1.08 &\ 0.97 &\ 1.99 \\\hline
    \ 1.38 &\ 1.24 &\ 2.57 \\
  \end{tabular} & \begin{tabular}{p{2em}|p{2em}|p{2em}} 
    \ 0.53 &\ 0.41 &\ 0.85 \\\hline
    \ 0.59 &\ 0.49 &\ 1.02 \\\hline
    \ 0.67 &\ 0.58 &\ 1.20 \\\hline
    \ 0.77 &\ 0.68 &\ 1.40 \\\hline
    \ 1.00 &\ 0.89 &\ 1.85 \\\hline
    \ 1.33 &\ 1.22 &\ 2.51 \\
  \end{tabular} & \begin{tabular}{p{2em}|p{2em}|p{2em}} 
    \ 0.59 &\ 0.47 &\ 0.98 \\\hline
    \ 0.72 &\ 0.64 &\ 1.30 \\\hline
    \ 0.79 &\ 0.71 &\ 1.45 \\\hline
    \ 0.87 &\ 0.79 &\ 1.61 \\\hline
    \ 1.15 &\ 1.07 &\ 2.18 \\\hline
    \ 1.70 &\ 1.61 &\ 3.27 \\
  \end{tabular} \\
\hline \hline 
\raisebox{-3em}{\rotatebox{90}{UrQMD: $y>0$}}
&\begin{tabular}{r} 10 \\\hline 20\\\hline 30\\\hline 40 \\\hline 80 \\\hline 158 \\\end{tabular}&
                  \begin{tabular}{p{2em}|p{2em}|p{2em}} 
    \ 0.84 &\ 0.43 &\ 0.63 \\\hline
    \ 0.89 &\ 0.48 &\ 0.82 \\\hline
    \ 0.91 &\ 0.55 &\ 0.97 \\\hline
    \ 0.98 &\ 0.62 &\ 1.09 \\\hline
    \ 1.02 &\ 0.68 &\ 1.31 \\\hline
    \ 1.19 &\ 0.69 &\ 1.46 \\
  \end{tabular} & \begin{tabular}{p{2em}|p{2em}|p{2em}} 
    \ 0.69 &\ 0.56 &\ 0.80 \\\hline
    \ 0.74 &\ 0.63 &\ 0.96 \\\hline
    \ 0.82 &\ 0.67 &\ 1.10 \\\hline
    \ 0.83 &\ 0.74 &\ 1.20 \\\hline
    \ 0.96 &\ 0.83 &\ 1.46 \\\hline
    \ 1.08 &\ 0.96 &\ 1.74 \\
  \end{tabular} & \begin{tabular}{p{2em}|p{2em}|p{2em}} 
    \ 0.71 &\ 0.55 &\ 0.80 \\\hline
    \ 0.73 &\ 0.60 &\ 0.90 \\\hline
    \ 0.77 &\ 0.65 &\ 1.02 \\\hline
    \ 0.83 &\ 0.70 &\ 1.14 \\\hline
    \ 0.90 &\ 0.81 &\ 1.37 \\\hline
    \ 1.07 &\ 0.95 &\ 1.70 \\
  \end{tabular} & \begin{tabular}{p{2em}|p{2em}|p{2em}} 
    \ 0.67 &\ 0.53 &\ 0.77 \\\hline
    \ 0.74 &\ 0.56 &\ 0.87 \\\hline
    \ 0.73 &\ 0.63 &\ 0.96 \\\hline
    \ 0.77 &\ 0.67 &\ 1.05 \\\hline
    \ 0.85 &\ 0.78 &\ 1.27 \\\hline
    \ 1.02 &\ 0.95 &\ 1.66 \\
  \end{tabular} & \begin{tabular}{p{2em}|p{2em}|p{2em}} 
    \ 0.71 &\ 0.56 &\ 0.83 \\\hline
    \ 0.76 &\ 0.65 &\ 0.99 \\\hline
    \ 0.78 &\ 0.69 &\ 1.06 \\\hline
    \ 0.81 &\ 0.72 &\ 1.13 \\\hline
    \ 0.92 &\ 0.85 &\ 1.41 \\\hline
    \ 1.18 &\ 1.12 &\ 1.96 \\
  \end{tabular} \\ \hline\hline
\end{tabular}
\vspace{0.5cm} \caption{The HSD and UrQMD scaled variances
$\omega_-$, $\omega_+$, and $\omega_{ch}$ for the 1\% of most
central collisions selected by largest values of $N_P^{proj}$. The
numbers correspond to those presented in
Figs.~\ref{wi_1pc_f}--\ref{wi_1pc_y2}. } \label{table1}
\end{center}
\end{table}

Fig.~\ref{wi_b0_f} shows that both HSD and UrQMD predict a
monotonic dependence of the charge particle multiplicity with
energy. So, the hadronic `background' for the NA61 experiments is
expected to be a smooth monotonic function of beam energy.

Besides of differences in the realization of the string
fragmentation model in HSD and UrQMD 1.3 mentioned above (cf. Fig.
1), additional deviations can be attributed to different
initializations of the nuclei in both models. Indeed, the
event-by-event observables show a higher sensitivity to the
initial nucleon density distribution than the standard single
particle observables \cite{Drescher07}. A pilot study using UrQMD
shows different $\omega$ when applying different initialization shapes.
Due to this effects a systematic error of 20\% is attributed to
$\omega$.
Such a sensitivity of the A-dependence of $\omega_{ch}$ to the
details of the models indicates a necessity for further studies of
the initializations of the nuclei in transport model approaches.
This becomes important for the theoretical interpretation of
future experimental data on event-by-event fluctuations.

Note that the model of independent sources and Eq.~(\ref{WMod})
work for the multiplicity fluctuations simulated by the transport
models in full $4\pi$ acceptance but not for the acceptance in a
specific rapidity region. The results for inelastic p+p collisions
are identical in the projectile and target hemispheres. This is
not the case in the sample of 1\% most central $A+A$ collisions
selected by $N_P^{proj}$. The total number of nucleons
participating in $A+A$ collisions fluctuates.  These fluctuations
are not symmetric in forward-backward hemispheres: in the selected
1\% sample the number of target participants $N_P^{targ}$
fluctuates essentially stronger than that of $N_P^{proj}$. The
consequences of the asymmetry in an event selection depend on the
dynamics of A+A collision (see Ref.~\cite{MGMG} for details). The
HSD and UrQMD results in Figs.~\ref{wi_1pc_y} and~\ref{wi_1pc_y2}
clearly demonstrate larger values for all scaled variances,
$\omega_-$, $\omega_+$, and $\omega_{ch}$, for $y<0$ acceptance
than those for $y>0$ one. This is due to stronger target
participant fluctuations, $\omega_P^{targ}>\omega_P^{proj}$.

\section{Summary and Conclusions}

In summary, the event-by-event multiplicity fluctuations in
nucleus-nucleus collisions have been studied for different
energies and system sizes within the HSD and UrQMD v. 1.3
transport approaches.  Our present study is in full correspondence
to the future experimental program of the NA61 Collaboration at
the SPS. Thus we have considered C+C, S+S, In+In, and Pb+Pb
nuclear collisions from $E_{lab}$= 10,~ 20,~ 30,~ 40,~ 80,
158~AGeV.  The influence of participant number fluctuations on
hadron multiplicity fluctuations has been emphasized and studied
in detail. To make these `trivial' fluctuations smaller, one has
to consider the most central collisions. Indeed, one needs to make a very
rigid selection -- 1\% or smaller -- of the `most central' collision
events. In addition, one wants to compare the event-by-event
fluctuations in these `most central' collisions for heavy and for
light nuclei. Under these new requirements different centrality
selections are not equivalent to each other. As a consequence,
there is no universal geometrical selection by the impact
parameter. This is a new and serious problem for theoretical
models (e.g., for hydrodynamical models) in a precision
description of the event-by-event fluctuation data. The above
statements have been illustrated by the $b=0$ selection criterium
considered in our paper. For light nuclei even these `absolutely central'
geometrical collisions lead to rather large fluctuations of the
number of participants, essentially larger than in the  1\% most
central collisions selected by the largest values of the
projectile participants $N_P^{proj}$.

We have, futhermore, used the number of projectile participants to define the
centrality selection. This is the most promising way of the
centrality selection in  fixed target experiments. It also
corresponds to the experimental plans of the NA61 collaboration. We
have defined the 1\% most central collisions by selecting the
largest values of the projectile participants $N_P^{proj}$. The
multiplicity fluctuations calculated in these samples show a much
weaker dependence on the atomic mass  number A than for criterium $b=0$.
A monotonic energy dependence for the multiplicity fluctuations
are obtained in both the HSD and UrQMD transport models. The two
models demonstrate a similar qualitative behavior of the particle
number fluctuations. However, the UrQMD 1.3 results for the scaled
variances $\omega_-$, $\omega_+$, and $\omega_{ch}$ are
systematically larger than those obtained within  HSD. This is
mainly due to the corresponding inequalities for the scaled
variances $\omega_{ch}$ (see Fig.~1, {\it right}) for p+p
collisions in these models. Our study has demonstrated  a
sensitivity of the multiplicity fluctuations to some specific
details of the transport models.
Nevertheless, the present HSD and UrQMD results for the scaled
variances provide a general trend of their dependencies on A and
$E_{lab}$ and also indicate quantitatively the systematic
uncertainties.

We stress again, that HSD and UrQMD do not include explicitly a
phase transition to the QGP. The expected enhanced fluctuations -
attributed to the critical point and phase transition - can be
observed experimentally on top of a monotonic and smooth `hadronic
background'.  The most promising signature of the QCD critical
point would be an observation of a non-monotonic dependence of the
scaled variances on bombarding energy $E_{lab}$ for central A+A
collisions with fixed atomic mass  number.   In the fixed target SPS
experiments the centrality selection in A+A collisions is defined
by the number of the projectile participants. The measurements of
$\omega_-$, $\omega_+$, and $\omega_{ch}$ are then preferable in
the forward hemispheres. In this case the remaining small
fluctuations of the number of target participants in the 1\%  most
central collisions become even less important, as they contribute
mainly to the particle fluctuations in the backward hemisphere.
Our findings should be helpful for the optimal choice of collision
systems and collision energies for the experimental search of the
QCD critical point.

\vspace{0.5cm} {\bf Acknowledgments}

We like to thank V.V.~Begun, M. Bleicher, W.~Cassing, M.~Ga\'zdzicki,
W.~Greiner, M.~Hauer, and I.N.~Mishustin for useful discussions.

\newpage


\begin{thebibliography}{00}

\bibitem{rev1}
H.~Heiselberg, Phys. Rep. {\bf 351}, 161 (2001).

\bibitem{rev2}
S.~Jeon and V.~Koch, Review for Quark-Gluon Plasma 3, eds. R.C.
Hwa and X.-N. Wang, World Scientific, Singapore, 430-490 (2004)
[arXiv:hep-ph/0304012].

\bibitem{rev3}
T.K. Nayak, arXiv:0706.2708 [nucl-ex].

\bibitem{ood}
  M.~Gazdzicki, M.~I.~Gorenstein, and S.~Mrowczynski,
  Phys. Lett. B {\bf 585}, 115 (2004);
  M.~I.~Gorenstein, M.~Gazdzicki, and O.~S.~Zozulya,
 {\it ibid.} {\bf 585}, 237 (2004).


\bibitem{fluc2} I.N. Mishustin, Phys. Rev. Lett. {\bf 82}, 4779 (1999);
Nucl. Phys. A {\bf 681}, 56c (2001); H. Heiselberg and A.D.
Jackson, Phys. Rev. C {\bf 63}, 064904 (2001).

\bibitem{SRS} M.A.~Stephanov, K.~Rajagopal, and E.V.~Shuryak,
Phys. Rev. Lett. {\bf 81}, 4816 (1998); Phys. Rev. D {\bf 60},
114028 (1999); M.A.~Stephanov, Acta Phys. Polon. B {\bf 35}, 2939
(2004);

\bibitem{fluc-mult} S.V.~Afanasev {\it et al}., [NA49 Collaboration],
Phys. Rev. Lett. {\bf 86}, 1965 (2001); M.M.~Aggarwal {\it et
al}., [WA98 Collaboration], Phys. Rev. C {\bf 65}, 054912 (2002);
J.~Adams {\it et al}., [STAR Collaboration], {\it ibid.}
{\bf 68}, 044905 (2003);  C.~Roland {\it et al}., [NA49
Collaboration], J. Phys. G {\bf 30} S1381 (2004); Z.W.~Chai {\it
et al}., [PHOBOS Collaboration], J. Phys. Conf. Ser. {\bf 37}, 128
(2005); M.~Rybczynski {\it et al.}  [NA49 Collaboration], J.\
Phys.\ Conf.\ Ser.\  {\bf 5}, 74 (2005); C. Alt {\it t al.}, [NA49
Collaboration], Phys. Rev. C {\bf 75} (2007) 064904.

\bibitem{fluc-pT}
H.~Appelshauser {\it et al.}  [NA49 Collaboration], Phys.\ Lett.\
B {\bf 459}, 679 (1999); D.~Adamova  {\it et al}., [CERES
Collaboration], Nucl. Phys. A {\bf 727}, 97 (2003); T. Anticic
{\it et al}., [NA49 Collaboration], Phys. Rev. C {\bf 70}, 034902
(2004); S.S.~Adler {\it et al}., [PHENIX Collaboration], Phys.
Rev. Lett.  {\bf 93}, 092301 (2004); J.~Adams  {\it et al}., [STAR
Collaboration], Phys. Rev. C {\bf 71}, 064906 (2005).





\bibitem{CE} V.V. Begun {\it et al.},
Phys. Rev. C {\bf 70}, 034901 (2004);
{\it ibid.} {\bf 71}, 054904 (2005);
{\it ibid.} {\bf 74}, 044903 (2006);
M. Hauer, V.V. Begun, and M.I. Gorenstein,
arXiv:0706.3290 [nucl-th]; M. Hauer, arXiv:0710.3938 [nucl-th].


\bibitem{KGB1}%
V.P.~Konchakovski, {\it et al.}, Phys. Rev. C {\bf 73}, 034902
(2006);
%
\bibitem{KGB2}%
V.P.~Konchakovski, {\it et al.},
Phys. Rev. C {\bf 74}, 064911 (2006).
%

\bibitem{Becattini}
    F. Becattini, J. Manninen, and M. Gazdzicki,
    Phys. Rev. C {\bf 73} (2006) 044905.

%


\bibitem{MCE}
 V.V. Begun, M.~Ga'zdzicki, M.I.~Gorenstein,
M.~Hauer, V.P.~Konchakovski, and B.~Lungwitz,
Phys. Rev. C {\bf 76}, 024902 (2007).

\bibitem{KGB3}%
V.P.~Konchakovski, M.I.~Gorenstein, and E.L.~Bratkovskaya, Phys.
Lett. {\bf B651}, 114 (2007).

\bibitem{LB}
B. Lungwitz and M. Bleicher, Phys. Rev. {\bf C76}, 044904 (2007).

\bibitem{NA49}
  B.~Lungwitzt {\it et al.} [NA49 Collaboration],
  arXiv:nucl-ex/0610046;
 C. Alt {\it et al.}, arXiv:0712.3216.


\bibitem{phenix}
S.S.~Adled {\it et al.} [PHENIX Collaboration], Phys. Rev. C {\bf
71}, 034908 (2005); {\it ibid.} {\bf 71}, 049901 (2005);
J.~Mitchell [PHENIX Collaboration], J. Phys. Conf. Ser. {\bf 27},
88 (2005).

\bibitem{KGB4}%
V.P.~Konchakovski, M.I.~Gorenstein, and E.L.~Bratkovskaya, Phys.
Rev. C {\bf 76}, 031901(R) (2007).


\bibitem{NA61}%
 M.~Gazdzicki {\it et al.}  [NA61 Collaboration],
 PoS C {\bf POD2006}, 016 (2006)
 [arXiv:nucl-ex/0612007];
 N.~Antoniou {\it et al.}  [NA61 Collaboration],
 ``Study of hadron production in hadron nucleus and nucleus nucleus
 collisions at the CERN SPS,''
 CERN-SPSC-P-330 (2006).

\bibitem{HSD}
W. Ehehalt and W. Cassing, Nucl. Phys. A {\bf 602}, 449 (1996);
W. Cassing and E.L. Bratkovskaya, Phys. Rep. {\bf 308}, 65 (1999).

\bibitem{UrQMD}
S.A. Bass et al., Prog. Part. Nucl. Phys.{\bf 41}, 255 (1998);
M.~Bleicher {\it et al.},  J. Phys. G {\bf 25}, 1859 (1999).

\bibitem{Weber}
H.~Weber, {\it et al.}, Phys. Rev. {\bf C67}, 014904 (2003);
E.L.~Bratkovskaya, {\it et al.},
{\it ibid.} {\bf 67}, 054905 (2003);
 {\it ibid}, {\bf 69}, 054907 (2004);
  Prog. Part. Nucl. Phys. {\bf 53}, 225 (2004);
  Phys. Rev. Lett., {\bf 92}, 032302 (2004).

\bibitem{MGMG}
M.~Ga\'zdzicki and M.~Gorenstein, Phys. Lett. B{\bf 640}, 155
(2006).

\bibitem{BF}
W. Broniowski and W. Florkowski, Phys. Rev. C {\bf 65}, 024905
(2002).


\bibitem{Drescher07}
B. Mattos Tavares, H.-J. Drescher, and T. Kodama,
arXiv:hep-ph/0702224.


\end{thebibliography}
\end{document}